\documentclass[10pt]{article}
\usepackage{amsmath}
\usepackage{graphicx}
\usepackage{color}
\usepackage{orcidlink}
\usepackage{hyperref}
\hypersetup{colorlinks=true, linkcolor=blue, citecolor=blue, urlcolor=blue}
\usepackage{caption}
\usepackage{setspace}
\usepackage{multirow}
\usepackage{multicol}
\usepackage{amssymb}
\usepackage{subfigure}
\RequirePackage[numbers,sort&compress]{natbib}
\begin{document}
\onehalfspacing

\begin{center}
\LARGE{Thermodynamics and Joule-Thomson Expansion for Schwarzschild-AdS Black Holes with Cloud of Strings and Quintessential-like Fluid }
\par\end{center}

\vspace{0.3cm}

\begin{center}
{\bf Faizuddin Ahmed\orcidlink{0000-0003-2196-9622}}\footnote{\bf faizuddinahmed15@gmail.com}\\
{\it Department of Physics, Royal Global University, Guwahati, 781035, Assam, India}\\
\vspace{0.2cm}
{\bf Saeed Noori Gashti\orcidlink{0000-0001-7844-2640}}\footnote{\bf saeed.noorigashti70@gmail.com; sn.gashti@du.ac.ir}\\ 
{\it School of Physics, Damghan University, P. O. Box 3671641167, Damghan, Iran}\\
\vspace{0.2cm}
{\bf Behnam Pourhassan\orcidlink{0000-0003-1338-7083}}\footnote{\bf b.pourhassan@du.ac.ir}\\ 
{\it School of Physics, Damghan University, P. O. Box 3671641167, Damghan, Iran\\
Center for Theoretical Physics, Khazar University, 41 Mehseti Street, Baku, AZ1096, Azerbaijan
}\\
\vspace{0.2cm}
{\bf Abdelmalek Bouzenada\orcidlink{0000-0002-3363-980X}}\footnote{\bf abdelmalekbouzenada@gmail.com}\\ 
{\it Laboratory of Theoretical and Applied Physics, Echahid Cheikh Larbi Tebessi University 12001, Algeria}

\end{center}

\begin{abstract}
In this study, we explore the thermodynamic properties of a Schwarzschild black hole (BH) embedded in an anti-de Sitter (AdS) background, which is further coupled with a cloud of strings and surrounded by a quintessence-like fluid. Beginning with the formulation of BH mass in terms of the event horizon radius, we incorporate the concept of pressure as related to the AdS curvature radius within the framework of extended phase space thermodynamics. Using this setup, we derive key thermodynamic quantities, including the Gibbs free energy and internal energy, to characterize the energetic behavior of the black hole system. To assess the stability of the black hole, we compute the specific heat capacity and analyze how it is influenced by external parameters, such as the string cloud and the quintessence-like fluid. These geometric and matter fields are shown to significantly modify the thermal response of the BH. Furthermore, we examine the inversion temperature associated with the black hole and highlight its distinction from the standard Hawking temperature, providing deeper insight into the phase structure. Additionally, we investigate the Joule-Thomson expansion process and demonstrate how the aforementioned parameters affect this thermodynamic phenomenon, showing important aspects of BH cooling and heating behavior in an extended thermodynamic context.
\end{abstract}

\section{Introduction}\label{sec:1}

Black hole (BH) thermodynamics has evolved into a central and insightful field of gravitational physics, illustrating our information between geometry, quantum theory, and statistical mechanics, with many important results illustrated in the literature \cite{ref1,ref2,ref3,ref4,ref5,ref6,ref7}. Fundamental thermodynamic variables, including entropy and temperature, are intimately related to geometric properties such as the event horizon’s area and the surface gravity. Semiclassical investigations reveal that the emission of Hawking radiation inherently drives BHs toward thermal instability, as the temperature rises with decreasing horizon size, initiating a runaway heating process. Over the years, extensive studies have been devoted to understanding various aspects of BH thermodynamics, among which the question of thermal stability remains crucial. This stability, reflecting the system’s resilience to small perturbations in its thermodynamic parameter influence, is often tested through its phase structure, especially near critical point information, using a variety of theoretical frameworks \cite{ref8}. One of the most widely adopted methods for assessing BH stability is the computation of specific heat ($C_p$) since a positive heat capacity signifies a thermodynamically stable configuration \cite{ref9,ref10}. Also, specific heat ($C_p$) analysis serves as a powerful diagnostic tool for exploring phase transitions in BH systems, where the sign change or divergence of the heat capacity provides insight into the underlying critical behavior \cite{ref10,ref11,ref12,ref13,ref14}, showing and illustrating two principal types: transitions indicated by finite changes in heat capacity ($C_v$) and those marked by its divergence.

In the domain of BH thermodynamics, anti-de Sitter (AdS) BHs have received considerable attention due to their rich phase structures and deep analogies with conventional thermodynamic systems. A notable milestone in this field was the discovery of the Hawking-Page phase transition, which illustrates and defines a shift between a Schwarzschild-AdS BH and thermal AdS spacetime, which also plays a crucial role in the AdS/CFT correspondence \cite{ref5}. Later, Chamblin \textit{et al.} established a remarkable correspondence between the phase behavior of charged AdS BHs and that of a liquid-gas system in van der Waals fluids, particularly in the context of Reissner-Nordström-AdS (RN-AdS) BHs \cite{BHT2, BHT3}. This connection was significantly deepened with the formulation of the extended phase space approach, where the cosmological constant is interpreted as a thermodynamic pressure, leading to the identification of $P-V$ criticality \cite{ref8}. Within this framework, other phenomena such as reentrant phase transitions \cite{BHT5, BHT6} and the presence of a triple point \cite{BHT7} have been tested, illustrating the idea that AdS BH models can show behaviors analogous to ordinary thermodynamic systems. The breadth of developments in this subject is captured in recent reviews such as \cite{BHT8}. In this context, explain the cosmological constant properties, as illustrated this concepts in this study \cite{BHT9}, offering a method for extracting mechanical work from BHs. Also, this study was further extended through its application to the rankine cycle, envisioning BHs as efficient cosmic-scale power plants \cite{BHT10}. Up to now, thermodynamic properties of AdS BHs have been widely studied in the literature (see, Refs. \cite{BPD1,BPD2,BPD3,BPD4,BPD5,BPD6,BPD7,BPD8,BPD9,BPD10,FA1,FA2}).

Another important line of inquiry is the Joule-Thomson expansion, first tested for charged AdS BH models in \cite{BHT11}, which investigated inversion curves and critical behavior, revealing both parallels and distinctions with van der Waals systems. This study was subsequently generalized to Kerr-AdS black holes \cite{BHT12}, holographic superfluids \cite{BHT13}, and quintessence-modified Reissner-Nordström-AdS black holes \cite{BHT14}, expanding the thermodynamic understanding of black holes across diverse gravitational frameworks. One of the significant areas of research within the field of black hole thermodynamics is the investigation of phase transitions. This topic involves employing a variety of analytical and computational methods to explore and characterize the different types of phase transitions that black holes can undergo. In recent years, particular attention has been directed towards the thermodynamic topology of black holes, which has emerged as a highly influential and insightful approach for examining the nature and properties of these phase transitions across diverse BH systems. This novel perspective has contributed substantially to advancing our understanding of black hole thermodynamics and the underlying physical mechanisms driving such transitions. For a more study on thermodynamic topology, readers can see these references \cite{a20,20a,22a,23a,24a,25a,26a,40a,44i}.

By incorporating the cosmological constant into the thermodynamic framework of BHs, it becomes possible to analyze these objects in an extended phase space, opening the way to study thermodynamic laws, the weak cosmic censorship conjecture, BH heat engines, and the Joule-Thomson (J-T) expansion process. The first detailed investigation of the J-T effect in charged AdS BHs was conducted in \cite{BHT11}, where the process, analogous to the classical thermodynamic expansion of a gas from high to low pressure through a porous plug under constant enthalpy, was formulated in the gravitational context. Since this pioneering work, the method has been extended to a wide range of BH geometries, including $d$-dimensional charged AdS BHs \cite{JT3}, RN-AdS BHs in $f(R)$ gravity \cite{JT4}, AdS BHs with a global monopole \cite{JT5}, regular (Bardeen)-AdS BHs \cite{JT6}, charged AdS BHs in rainbow gravity \cite{JT7}, Hayward-AdS BHs \cite{JT8}, AdS BHs with momentum relaxation \cite{JT9}, Bardeen-AdS BH \cite{JT10,JT10a}, Lovelock gravity BHs \cite{JT11}, charged Gauss-Bonnet AdS BHs \cite{JT12}, AdS BHs in quasitopological electromagnetism \cite{JT12a}, charged Hayward-AdS BH in extended phase space \cite{JT12b}, and Hayward-AdS BH surrounded by fluid of strings \cite{JT12c}, Hayward-AdS BHs in 4D Einstein-Gauss-Bonnet gravity \cite{JT16}.

Our study focuses on the comprehensive thermodynamic analysis of a Schwarzschild BH embedded in an AdS spacetime, which is further modified by the presence of a cloud of strings and a surrounding quintessence-like fluid. Within the framework of extended phase space thermodynamics, where the negative cosmological constant is reinterpreted as a thermodynamic pressure related to the AdS curvature radius, we start by formulating the BH mass as a function of the event horizon radius. In this case, this formulation enables us to derive and examine a range of fundamental thermodynamic quantities, including the Gibbs free energy, internal energy, and specific heat capacity, providing a detailed description of the energetic and stability properties of the system. Also, our results show and illustrate the combined effects of the string cloud parameter and the quintessence-like fluid introduce significant modifications to the BH’s thermal behavior, altering its heat capacity profile and stability domains. Also, to further illustrate the phase structure, we compute the inversion temperature and compare it with the Hawking temperature, identifying distinct features that shed light on the role of external fields in BH thermodynamics. In addition, we investigate the Joule–Thomson expansion process, emphasizing how the interplay between the cloud of strings and the quintessence-like background affects the transition between heating and cooling regimes, as well as the location of the inversion curves. In this case, this extended thermodynamic perspective not only deepens our understanding of BH systems with additional matter fields but also shows the intricate ways in which geometric modifications and exotic fluids influence both local stability and global phase behavior.

The structure of this paper is organized as follows: In Section \ref{sec:1}, we present the introduction to our study. Also, Section \ref{sec:2} is devoted to the thermodynamic properties of the considered BH configuration, where we derive key quantities and explore their dependence on the system’s parameters. In this case, in Section \ref{S3}, we investigate the stability of the BH by testing specific heat behavior and related thermodynamic criteria. Another important result, Section \ref{S4} focuses on the thermodynamic geometry, employing geometric methods to illustrate the phase structures and BH parameter interactions. In this context, in Section \ref{S5}, we determine the inversion temperature, which marks the transition between cooling and heating phases under the Joule–Thomson process. The next section \ref{S6}, is a detailed discussion of the Joule–Thomson expansion, where we analyze the isenthalpic curves and the corresponding physical implications. Finally, Section \ref{S7} shows the main findings, illustrates the significance of the results, and suggests possible directions for future research. 

\section{BH Thermodynamics}\label{sec:2}

In this section, we explore the thermodynamic properties of an AdS BH model coupled to a cloud of strings and surrounded by a quintessence-like fluid. We derive the corresponding thermodynamic quantities-such as the Hawking temperature, Gibbs free energy, internal energy, and specific heat capacity and examine how the presence of CoS and QF affects these variables. In this framework, the BH mass is interpreted as the enthalpy, the event horizon area as the entropy \cite{DK}, and the cosmological constant is identified with the thermodynamic pressure in extended phase space formalism \cite{ref5,ref8}.

In this work, we are main interested on the following static and spherically symmetric AdS space-time with a CoS and surrounded by a QF described by the line element \cite{arxiv,AK}
\begin{equation}
ds^2 = -f(r)\,dt^2 +\frac{1}{f(r)}\,dr^2 + r^2\,(d\theta^2 + \sin^2\theta d\phi^2),\label{final}
\end{equation}
with the metric function is given by
\begin{equation}
f(r) =1 - \frac{2\,M}{r} + \frac{\lvert \alpha\rvert\, b^2}{r^2}\,{}_2F_1\left(-\frac{1}{2},-\frac{1}{4},\frac{3}{4},-\frac{r^4}{b^4}\right)-\frac{\mathrm{N}}{r^{3\,w+1}}+\frac{r^2}{\ell^2_p}\quad\quad (-1 < w <-1/3),\label{function}
\end{equation}
Here $(|\alpha|,b)$ represents the CoS parameters \cite{PSL,GA} and $(\mathrm{N}, w)$ denotes the QF parameters \cite{VVK}.

Let us now proceed to study this process for the BH. First of all, let us compute the BH mass \(M\), which is defined as a function of the location of its horizon \(r_+\), through the largest root of \(f(r=r_{+})=0\), before any cosmological horizon. Then by using Eq.~(\ref{function}), one gets
\begin{equation}
M = \frac{1}{2}\,\left[r_{+}+ \frac{\lvert \alpha\rvert\, b^2}{r_{+}}\,{}_2F_1\left(-\frac{1}{2},-\frac{1}{4},\frac{3}{4},-\frac{r^4_{+}}{b^4}\right)-\frac{\mathrm{N}}{r^{3\,w}_{+}}+\frac{8\pi}{3}\,P\,r^3_{+}\right],\label{bb1}
\end{equation}
with \(M\) interpreted as the enthalpy of the system and we have used the thermodynamic pressure in extended phase space of the following form \cite{ref8}:
\begin{equation}
P =\frac{3}{8\,\pi\,\ell^2_p}.\label{dd2}
\end{equation}
The Thermodynamic volume $V$ can be determined useing the following relation:
\begin{equation}
V=\frac{\partial M}{\partial P}=\frac{4\pi}{3}\,r^3_{+}.\label{dd2a}
\end{equation}

The Hawking temperature is given by
\begin{align}\label{bb2}
T =\frac{f'(r_{+})}{4\pi}=\frac{1}{4\pi\,r_{+}}\,\Bigg[1-|\alpha|\,b^2\,\left\{\frac{1}{r_+^2}\,{}_2F_1\left(-\tfrac{1}{2}, -\tfrac{1}{4}, \tfrac{3}{4}, -\tfrac{r_+^4}{b^4} \right)+ \frac{2\, r^2_+}{3\,b^4}\, {}_2F_1\left(\tfrac{1}{2}, \tfrac{3}{4}, \tfrac{7}{4}, -\tfrac{r_+^4}{b^4} \right)\right\}+3\,w\,\mathrm{N}\,r_+^{-3w-2}+8\,\pi\,P\,r^2_+\Bigg].
\end{align}

From expression (\ref{bb2}), we observe that the Hawking temperature is modified by the CoS parameters $(\alpha,\,b)$, the QF parameters \((\mathrm{N},\,w)\), and the thermodynamic pressure $P$.  

\begin{figure}[ht!]
    \centering
    \includegraphics[width=0.45\linewidth]{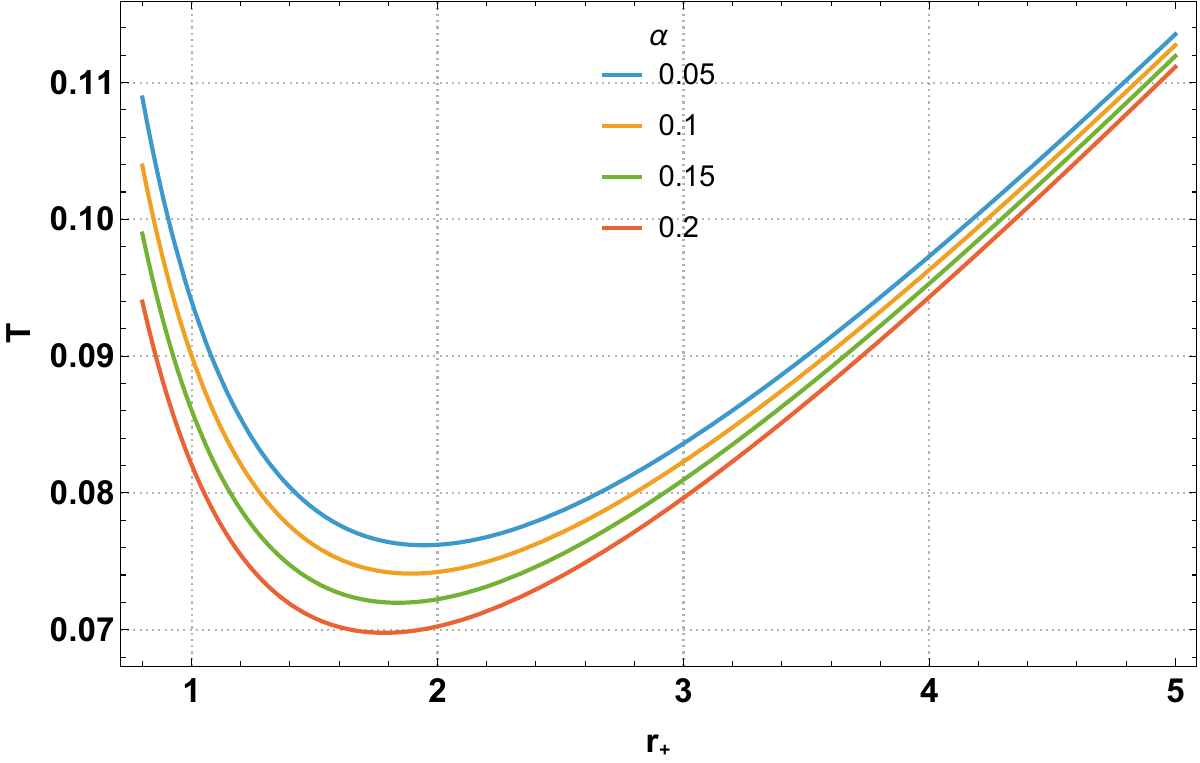}\quad\quad
    \includegraphics[width=0.50\linewidth]{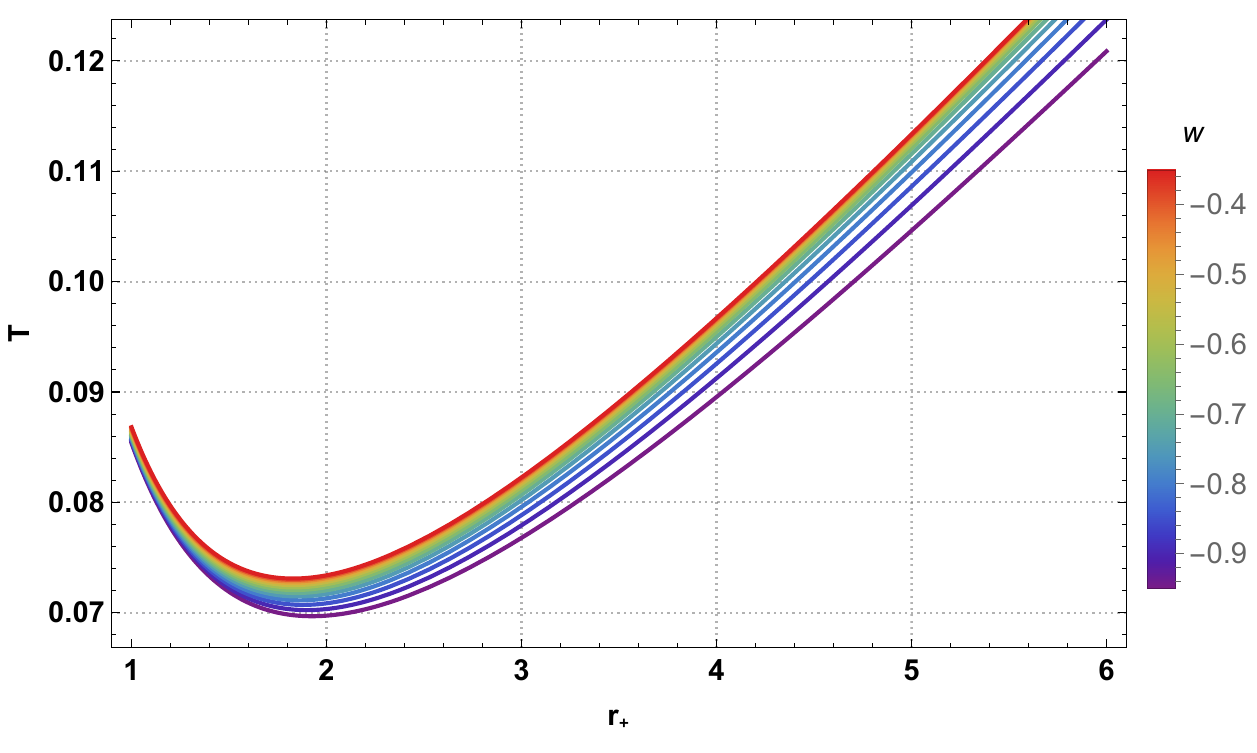}\\
     (i) $b=0.1,\,w=-2/3,\,\mathrm{N}=0.01,\,P=0.01$ \hspace{3cm} (ii) $\alpha=0.15,\,b=0.1,\,\mathrm{N}=0.01,\,P=0.01$
    \caption{\footnotesize Behavior of the Hawking Temperature $T$ given in Eq. (\ref{bb2}) as a function of horizon for different values of CoS parameter $\alpha$ and state parameter $W$.}
    \label{fig:Hawking-temperature}
\end{figure}
 
Fig. (\ref{fig:Hawking-temperature}) illustrates the behavior of the Hawking temperature as a function of the event horizon radius $r_+$ for Schwarzschild–AdS black holes immersed in a cloud of strings and surrounded by a quintessential-like fluid. In these plots, we consider some values for the free parameters—such as the CoS and QF parameters. Also, we set equation of state parameter $\omega$, $N$ and $P$—are held fixed within each subfigure to isolate their individual effects. The zero-temperature points coincide with the divergence points of the Joule-Thomson coefficient, indicating critical thresholds in the black hole’s thermodynamic phase space. This correlation underscores the deep connection between the black hole’s temperature profile and its cooling/heating behavior during expansion or compression processes. Moreover, the thermal behavior exhibits characteristics depending on the size of the event horizon radius. Here for all horizon radii, the temperature profiles show markedly same structures reflecting the sensitivity to the black hole’s microphysical parameters and the surrounding matter fields. 

We can more prominently analyze the behavior of the Gibbs free energy, $F$, of this black hole, as it provides valuable insights into the thermodynamic processes occurring within the black hole system, particularly those related to phase transitions and critical phenomena. The Gibbs free energy, a fundamental thermodynamic potential, is defined by the relation;
\begin{equation}\label{F1}
F = M - T\, S,
\end{equation}
where $M$ denotes the mass (or internal energy) of the black hole, $T$ represents its Hawking temperature, and $S$ is the entropy associated with the black hole horizon. 

To compute the entropy of the BH, one can see in Ref. \cite{MS} that this entropy is the same as Hawking-Bekenstein entropy, so that
\begin{equation}
S = \pi\,r_+^2.\label{bb3}
\end{equation}
This definition encapsulates the balance between the energy content and thermal effects of the black hole system, making $F$ a crucial quantity for studying its stability and phase behavior. By substituting the explicit expressions for $M$, $T$, and $S$, as previously derived in Eqs. (\ref{bb1}), (\ref{bb2}), and (\ref{bb3}) respectively, one can obtain an explicit analytical form for the Gibbs free energy:
\begin{align}\label{F2}
F =\frac{1}{4} r_+ 
+ \frac{|\alpha|\, b^2}{4\, r_+}\,{}_2F_1\left(-\tfrac{1}{2}, -\tfrac{1}{4}, \tfrac{3}{4}, -\tfrac{r_+^4}{b^4} \right)- \frac{|\alpha|\, r_+^3}{6\, b^2}\,{}_2F_1\left(\tfrac{1}{2}, \tfrac{3}{4}, \tfrac{7}{4}, -\tfrac{r_+^4}{b^4} \right)- \left( \frac{2 + 3w}{4} \right)\,\frac{\mathrm{N}}{r_+^{3w}}-\frac{2\,\pi}{3}\,P\,r_+^3
\end{align}

\begin{figure}[ht!]
    \centering
    \includegraphics[width=0.465\linewidth]{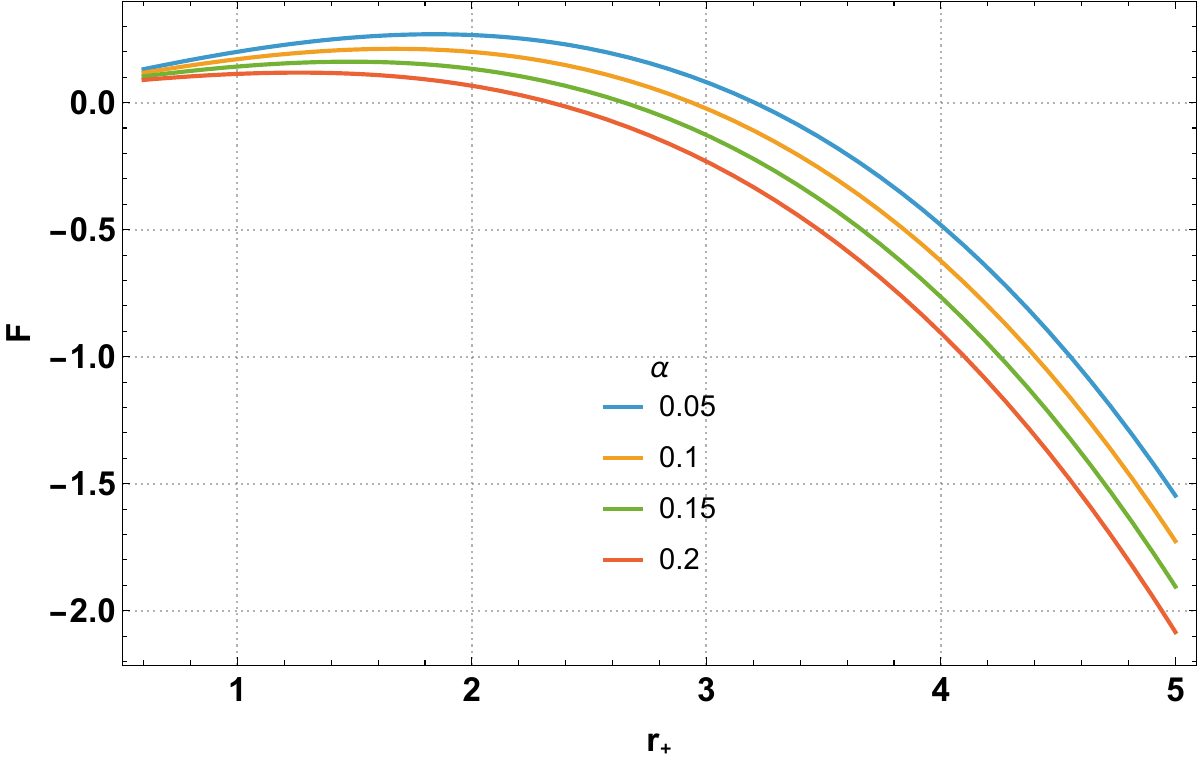}\quad
    \includegraphics[width=0.5\linewidth]{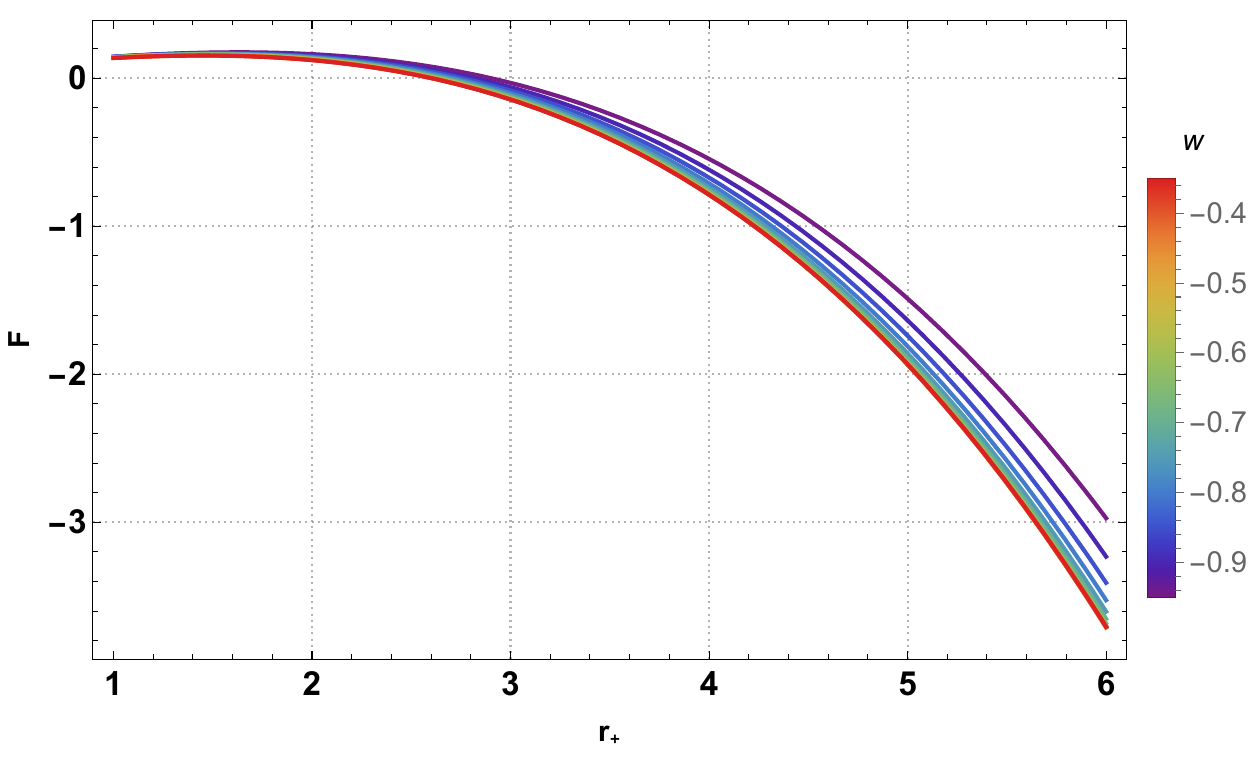}\\
    (i) $b=0.2,\,w=-2/3$ \hspace{6cm} (ii) $\alpha=0.1,\,b=0.2$
    \caption{\footnotesize Behavior of the Gibb's free energy $F$ given in Eq. (\ref{F2}) as a function of horizon for different values of CoS parameter $\alpha$ and state parameter $W$. Here, $\mathrm{N}=0.01,\,P=0.01$.}
    \label{fig:Free-energy}
\end{figure}

From expression (\ref{F2}), we observe that the Gibb's free energy is modified by the CoS parameters $(\alpha,\,b)$, the QF parameters \((\mathrm{N},\,w)\), and the thermodynamic pressure $P$.

Finally, the internal energy of the system is given by
\begin{equation}
    U=M-P\,V=\frac{1}{2}\,\left[r_{+}+ \frac{\lvert \alpha\rvert\, b^2}{r_{+}}\,{}_2F_1\left(-\frac{1}{2},-\frac{1}{4},\frac{3}{4},-\frac{r^4_{+}}{b^4}\right)-\frac{\mathrm{N}}{r^{3\,w}_{+}} \right].\label{F3}
\end{equation}

\begin{figure}[ht!]
    \centering
    \includegraphics[width=0.46\linewidth]{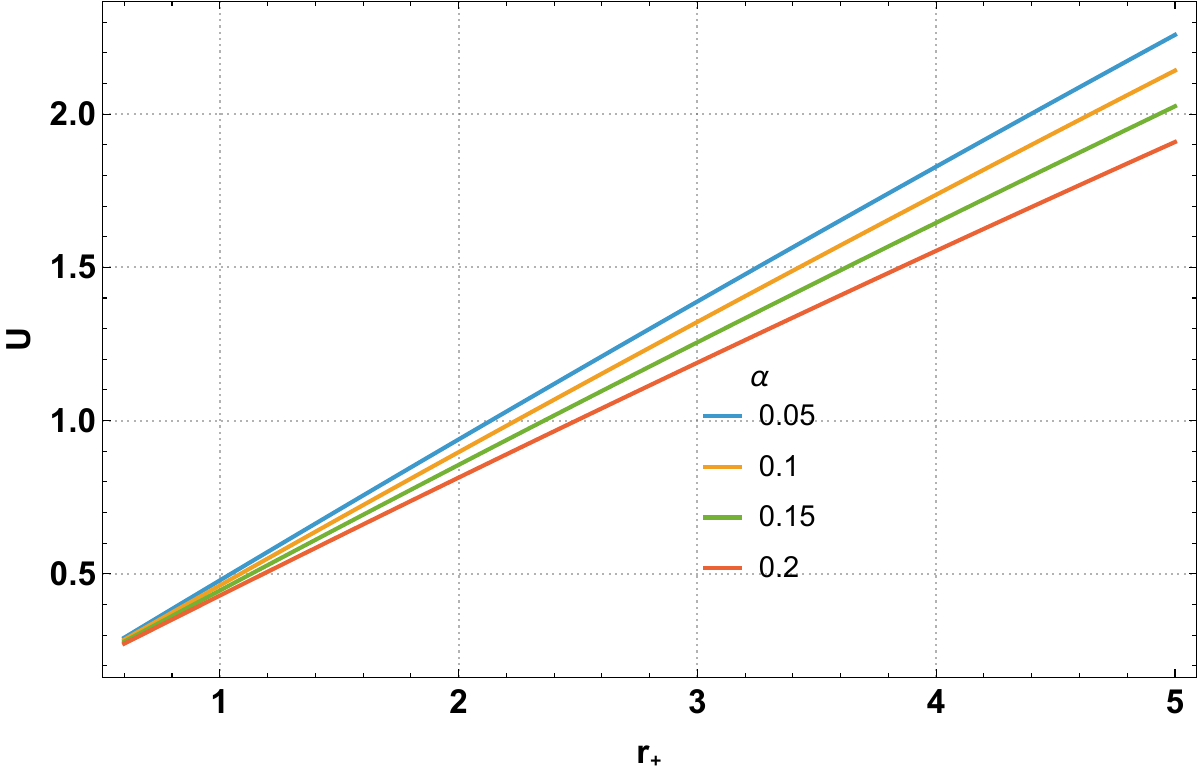}\quad
    \includegraphics[width=0.5\linewidth]{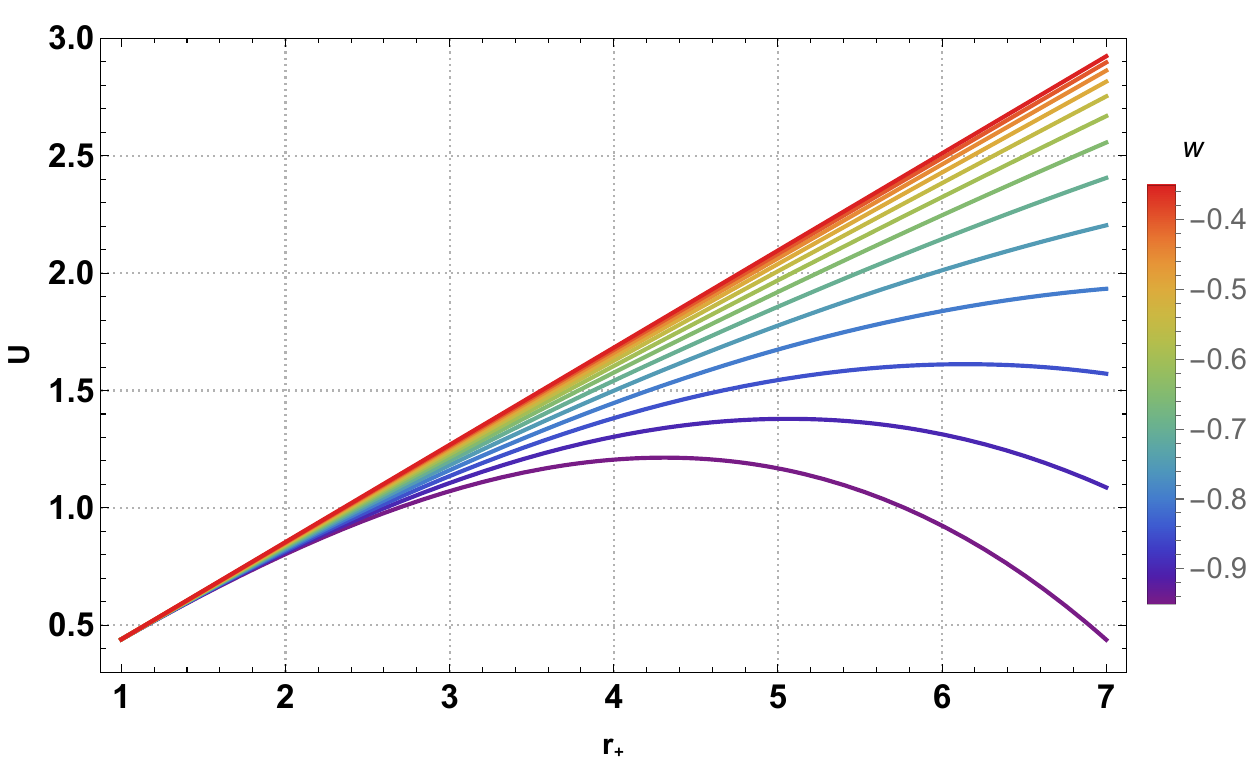}\\
    (i) $b=0.2,\,w=-2/3,\,\mathrm{N}=0.01$ \hspace{6cm} (ii) $\alpha=0.1,\,b=0.2,\,\mathrm{N}=0.01$
    \caption{\footnotesize Behavior of the internal energy $U$ given in Eq. (\ref{F3}) as a function of horizon for different values of CoS parameter $\alpha$ and state parameter $W$.}
    \label{fig:internal-energy}
\end{figure}

From expression (\ref{F3}), we observe that the internal energy is modified by the CoS parameters $(\alpha,\,b)$, the QF parameters \((\mathrm{N},\,w)\), and the thermodynamic pressure $P$.

Eq. (\ref{F2}) explicitly relates $F$ to the horizon radius $r_+$ and other relevant parameters of the black hole, enabling a detailed analysis of its thermodynamic properties. Fig. \ref{fig:Free-energy} presents the variation of the free energy $F$ as a function of the horizon radius $r_+$. From this plot, it is evident that the free energy exhibits distinct features at $r_+$. These features signify the occurrence of phase transitions, indicating changes in the stability and configuration of the black hole states. The behavior of $F$ with respect to $r_+$ thus serves as a powerful diagnostic tool for understanding the thermodynamic phase structure of the black holes under consideration.
In particular, the presence of multiple branches or turning points in the $F$ versus $r_+$ curve suggests rich thermodynamic phenomena, such as first-order or second-order phase transitions, and critical points. Such phenomena are essential for comprehending the microphysical processes governing black hole thermodynamics and can shed light on the interplay between gravity, quantum effects, and thermodynamics in these extreme systems.

Furthermore, as illustrated in Fig.~~(\ref{fig:internal-energy}), we analyze the behavior of the internal energy $U$, as defined in Eq.~~(\ref{F3}), plotted as a function of the black hole horizon radius. This analysis is carried out for various values of the coefficient of state (CoS) parameter $\alpha$ as well as different values of the equation of state parameter $\omega$. The figure clearly demonstrates how changes in these parameters influence the thermodynamic behavior of the system, particularly highlighting the sensitivity of the internal energy to both the gravitational modifications introduced by $\alpha$ and the matter content encoded in $\omega$. Such dependencies offer deeper insight into the interplay between geometry and matter in modified gravity scenarios and can shed light on the underlying microphysics near black hole horizons.

\subsection{Specific Heat Capacity: Stability of BH}\label{S3}

To comprehensively understand the thermodynamic behavior and stability characteristics of the Schwarzschild-AdS black holes with cloud of strings and quintessential-like fluid, we analyze its specific heat at constant parameters. This quantity provides critical insights into the response of the system to thermal fluctuations and serves as an indicator of local thermodynamic stability. The specific heat is derived using the expressions for the Hawking temperature $T$, given in Eq.~~(\ref{bb2}), and the entropy $S$, given in Eq.~~(\ref{bb3}). Employing the fundamental thermodynamic identity, the specific heat at constant parameters is defined as \cite{ref8,SWW}:
\begin{equation}\label{C1}
C_p = T\,\left( \frac{\partial S}{\partial T} \right) = T\,\left( \frac{\partial S}{\partial r_{+}} \right)\,\left( \frac{\partial T}{\partial r_{+}}\right)^{-1},
\end{equation}

Using the Hawking temperature given in Eq.~(\ref{bb2}) and the entropy in Eq. (\ref{bb3}), the specific heat capacity simplifies to {\footnotesize
\begin{align}\label{C3}
C_p=\frac{2\,\pi\,r^2_{+}\,\left[\frac{1}{r^2_{+}}-\frac{|\alpha|\,b^2}{r_{+}}\,\left\{\frac{1}{r_+^3}\,{}_2F_1\left(-\tfrac{1}{2}, -\tfrac{1}{4}, \tfrac{3}{4}, -\tfrac{r_+^4}{b^4} \right)
+ \frac{2}{3\,b^4}\, {}_2F_1\left(\tfrac{1}{2}, \tfrac{3}{4}, \tfrac{7}{4}, -\tfrac{r_+^4}{b^4} \right)\right\}+3\,w\,\mathrm{N}\,r_+^{-3w-3}+8\,\pi\,P\right]}{\left[-\frac{1}{r_+^2}
+ |\alpha|\, b^2\, \left\{\frac{3}{r_+^4}\,{}_2F_1\left(-\tfrac{1}{2}, -\tfrac{1}{4}, \tfrac{3}{4}, -\tfrac{r_+^4}{b^4} \right)+ \frac{2}{3\,b^4}\,{}_2F_1\left(\tfrac{1}{2}, \tfrac{3}{4}, \tfrac{7}{4}, -\tfrac{r_+^4}{b^4} \right) - \frac{6,r_+^3}{7\,b^8}\,{}_2F_1\left(\tfrac{3}{2}, \tfrac{7}{4}, \tfrac{11}{4}, -\tfrac{r_+^4}{b^4} \right)\right\}-\frac{3w\,\mathrm{N} (3w + 2)}{r_+^{3w + 3}}+ 8\,\pi\, P\right]}
\end{align}
}

From expression (\ref{C3}), we observe that the specific heat capacity is modified by the CoS parameters $(\alpha,\,b)$ and the QF parameters \((\mathrm{N},\,w)\).

For a particular state parameter, $w=-2/3$, the specific heat reduces as,
\begin{align}\label{C4}
C_p=\frac{2\,\pi\,r^2_{+}\,\left[\frac{1}{r^2_{+}}-\frac{|\alpha|\,b^2}{r_{+}}\,\left\{\frac{1}{r_+^3}\,{}_2F_1\left(-\tfrac{1}{2}, -\tfrac{1}{4}, \tfrac{3}{4}, -\tfrac{r_+^4}{b^4} \right)
+ \frac{2}{3\,b^4}\, {}_2F_1\left(\tfrac{1}{2}, \tfrac{3}{4}, \tfrac{7}{4}, -\tfrac{r_+^4}{b^4} \right)\right\}-\frac{2\,\mathrm{N}}{r_+}+8\,\pi\,P\right]}{\left[-\frac{1}{r_+^2}
+ |\alpha|\, b^2\, \left\{\frac{3}{r_+^4}\,{}_2F_1\left(-\tfrac{1}{2}, -\tfrac{1}{4}, \tfrac{3}{4}, -\tfrac{r_+^4}{b^4} \right)+ \frac{2}{3\,b^4}\,{}_2F_1\left(\tfrac{1}{2}, \tfrac{3}{4}, \tfrac{7}{4}, -\tfrac{r_+^4}{b^4} \right) - \frac{6\,r_+^3}{7\,b^8}\,{}_2F_1\left(\tfrac{3}{2}, \tfrac{7}{4}, \tfrac{11}{4}, -\tfrac{r_+^4}{b^4} \right)\right\}+ 8\,\pi\, P\right]}
\end{align}

\begin{figure}[ht!]
    \centering
    \includegraphics[width=0.48\linewidth]{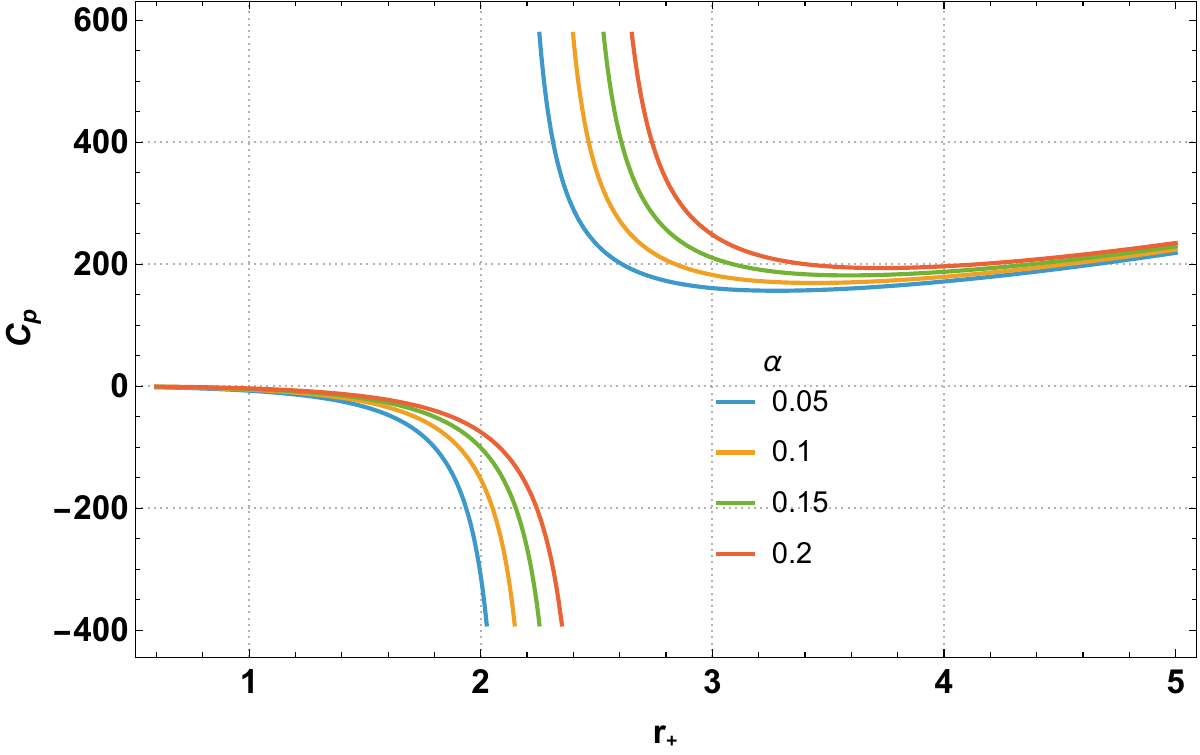}\quad
    \includegraphics[width=0.48\linewidth]{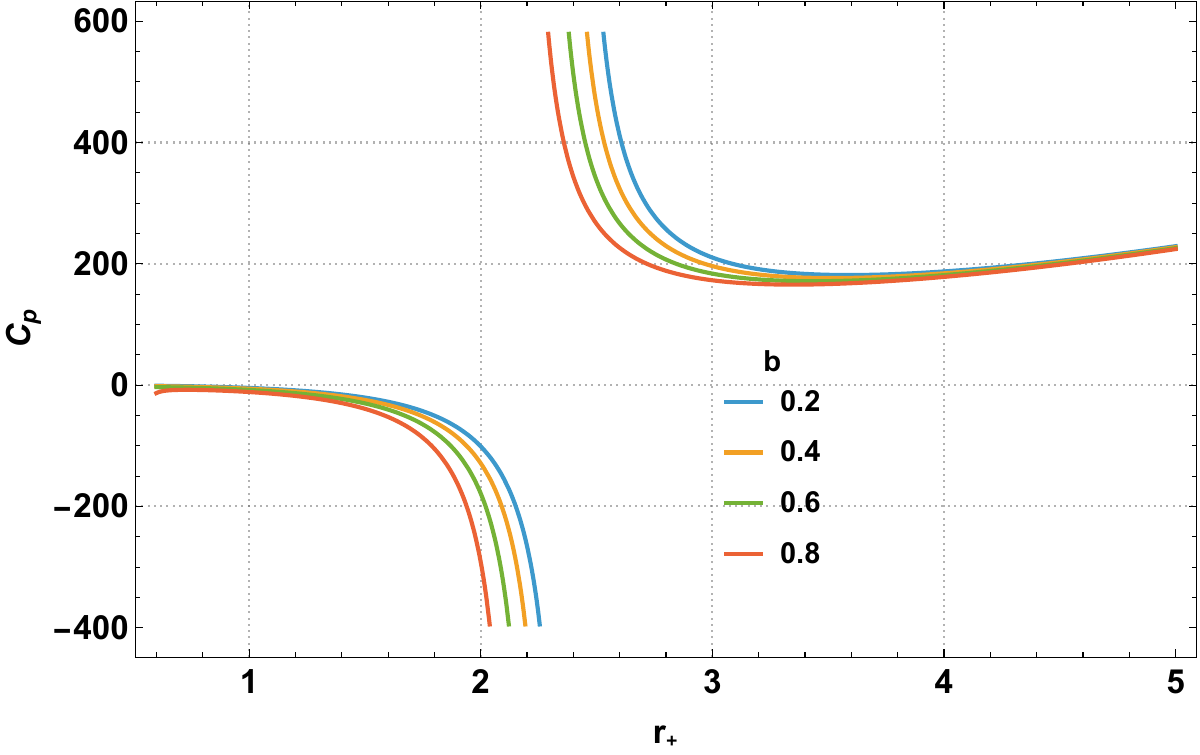}\\
    (i) $b=0.2$ \hspace{8cm} (ii) $\alpha=0.1$ 
    \caption{\footnotesize Behavior of the specific heat capacity $C_p$ given in Eq. (\ref{C4}) as a function of horizon for different values of CoS parameters $\alpha$ and $b$. Here, we set $w=-2/3,\,\mathrm{N}=0.01,\,P=0.01$.}
    \label{fig:heat-capacity}
\end{figure}

The analysis of the specific heat not only identifies the conditions under which the BH undergoes phase transitions but also provides a consistent thermodynamic interpretation of stability properties inferred from topological methods. The interplay between thermal stability and topological characteristics underscores the robustness of the thermodynamic framework in exploring modified gravity scenarios.

The behavior of $C_p$ as a function of the black hole parameters reveals several notable features that determine in Fig. (\ref{fig:heat-capacity}):

\begin{itemize}
\item \textbf{Phase Transitions:} The specific heat diverges at certain critical values of the horizon radius or other control parameters. These divergences signal second-order phase transitions, marking qualitative changes in the thermodynamic structure of the black hole.

\item \textbf{Thermal Stability:} In regions where the specific heat is positive (\( C_p > 0 \)), the black hole exhibits thermal stability, indicating that it can maintain equilibrium under small perturbations. Conversely, negative specific heat (\( C_p < 0 \)) corresponds to thermal instability, a characteristic feature of many black hole systems.

The analysis of the specific heat not only identifies the conditions under which the black hole undergoes phase transitions but also provides a consistent thermodynamic interpretation of stability properties inferred from topological methods. The interplay between thermal stability and topological characteristics underscores the robustness of the thermodynamic framework in exploring modified gravity scenarios.
\end{itemize}

\section{Thermodynamic Geometry of AdS BH with CoS and QF}\label{S4}

In addition to the standard thermodynamic stability analysis, it is instructive to examine the Schwarzschild–AdS black hole with a cloud of strings and quintessential-like fluid from the perspective of thermodynamic geometry. This approach, known as geometrothermodynamics \cite{G1, G2}, equips the space of equilibrium states with a Riemannian metric whose curvature encodes information about the microscopic interactions underlying the thermodynamic system \cite{G3, G4}.
In this framework, we adopt the Ruppeiner metric, which is defined in the entropy representation as,
\begin{equation}
	g^{R}_{ij} = -\frac{\partial^2 S}{\partial X^i \partial X^j},
\end{equation}
where $X^i$ are the extensive variables of the system. Equivalently, the Ruppeiner metric is conformally related to the Weinhold metric via $g^{R}_{ij} = \frac{1}{T} g^{W}_{ij}$, with
\begin{equation}
	g^{W}_{ij} = \frac{\partial^2 M}{\partial X^i \partial X^j}.
\end{equation}
For our black hole configuration, we choose the thermodynamic coordinates $X^i = (S, P)$ while holding $(|\alpha|, b, N, w)$ fixed. Using Eq.~(\ref{bb1}), the Weinhold metric components are
\begin{align}
	g^{W}_{SS} &= \frac{\partial^2 M}{\partial S^2}, \\
	g^{W}_{SP} &= \frac{\partial^2 M}{\partial S \partial P}=g^{W}_{PS}, \\
	g^{W}_{PP} &= \frac{\partial^2 M}{\partial P^2},
\end{align}
where
\begin{equation}
	M(S,P) = \frac{\sqrt{\pi}}{2} \left[
	\sqrt{S} 
	+ \frac{\pi |\alpha| b^{2} \sqrt{S}}{\;2F_1\!\left(-\frac{1}{2}, -\frac{1}{4}; \frac{3}{4}; -\frac{S^{2}}{\pi^{2} b^{4}}\right)}
	- N S^{- \frac{3w}{2}} \, \pi^{\frac{3w+1}{2}}
	+ \frac{8P}{3} S^{3/2}
	\right],
	\label{eq:MSP}
\end{equation}

The Ruppeiner curvature scalar $R_{R}$, obtained from the thermodynamic metric $g^{R}_{ij}$, encodes valuable information about the microscopic interactions of the system. A positive value of $R_{R}$ corresponds to predominantly repulsive interactions between the hypothetical microscopic constituents of the black hole, whereas a negative value signals the dominance of attractive interactions. When $R_{R}$ diverges, it typically marks the location of a phase transition, coinciding with points where other thermodynamic quantities such as the specific heat also become singular. Thus, the sign and divergence structure of $R_{R}$ provide a geometric probe of the black hole’s thermodynamic stability and microphysical behavior.
By explicitly computing $R_{R}$ for various values of $(\alpha, b, N, w)$, we find,

\begin{eqnarray}
	R_{R}(S,P) &=&
	\frac{
		-\frac{\sqrt{\pi}}{8} S^{-5/2}
		+ \frac{3w(3w+2)}{8} N \pi^{\frac{3w+1}{2}} S^{- \frac{3w}{2} - 2}
	}{
		\frac{32\pi}{9} \, T(S,P) \, S
	}\nonumber\\
	&+&\frac{
		9 \sqrt{\pi} |\alpha| b^{2} \left[
		\frac{15}{8} S^{-7/2} \, {}_{2}F_{1}\!\left(-\frac12,-\frac14;\frac34;-\frac{S^{2}}{\pi^{2} b^{4}}\right)
		+ \frac{5 S^{-3/2}}{3\pi^{2} b^{4}} \, {}_{2}F_{1}\!\left(\frac12,\frac34;\frac74;-\frac{S^{2}}{\pi^{2} b^{4}}\right)
		+ \frac{2 S^{1/2}}{9\pi^{4} b^{8}} \, {}_{2}F_{1}\!\left(\frac32,\frac74;\frac{11}{4};-\frac{S^{2}}{\pi^{2} b^{4}}\right)
		\right]
	}{
		64 \, T(S,P) \, S
	}.
	\label{eq:RRfinal}
\end{eqnarray}

where
\begin{equation}
	T(S,P) = \frac{1}{4 \sqrt{\pi}} \left[
	\frac{1}{\sqrt{S}}
	- \pi |\alpha| b^{2} \left\{
	\frac{1}{S^{3/2}} \, {}_{2}F_{1}\!\left(-\frac{1}{2}, -\frac{1}{4}; \frac{3}{4}; -\frac{S^{2}}{\pi^{2} b^{4}}\right)
	+ \frac{2\sqrt{S}}{3\pi^{2} b^{4}} \, {}_{2}F_{1}\!\left(\frac{1}{2}, \frac{3}{4}; \frac{7}{4}; -\frac{S^{2}}{\pi^{2} b^{4}}\right)
	\right\}
	+ 3w N S^{- \frac{3w}{2} - 1} \pi^{\frac{3w+1}{2}}
	+ 8P \sqrt{S}
	\right].
	\label{eq:TSP}
\end{equation}

\begin{figure}[ht!]
	\centering
	\includegraphics[width=0.45\textwidth]{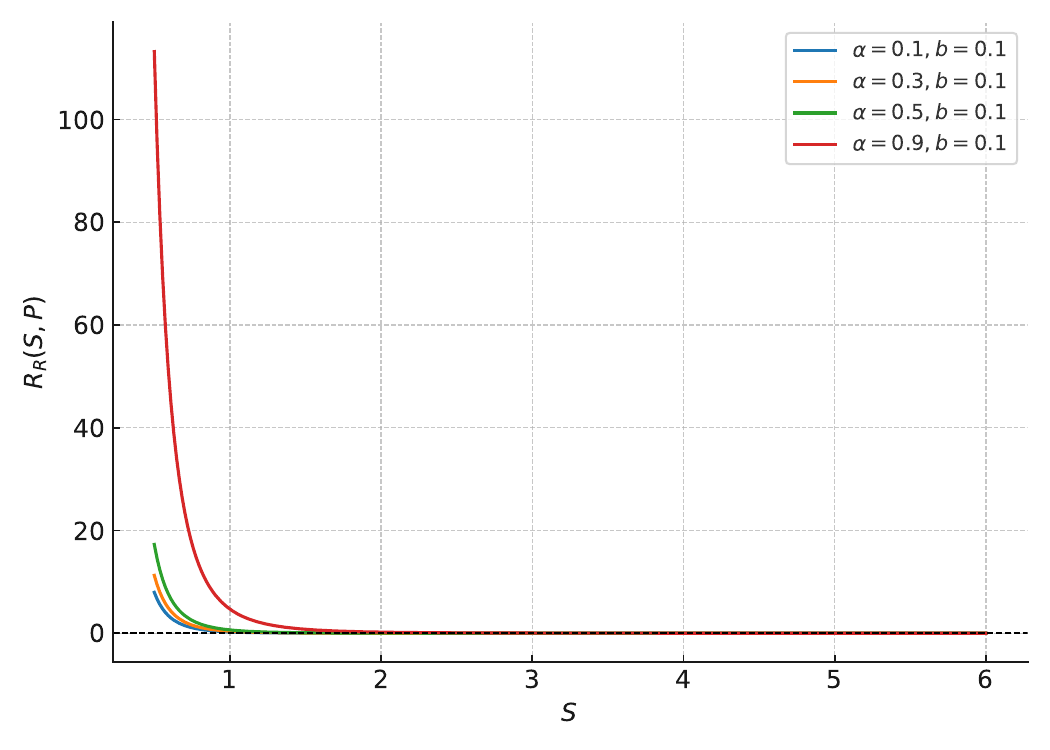}\quad\quad\quad
    \includegraphics[width=0.45\textwidth]{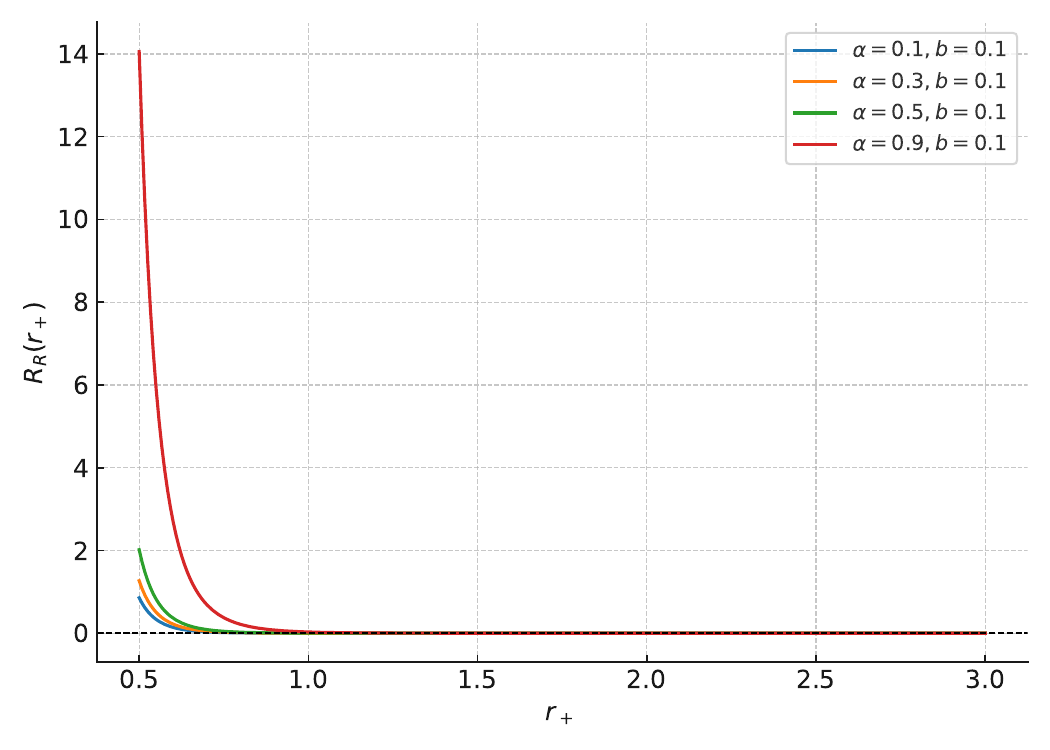}
	\caption{Ruppeiner curvature scalar $R_R$ as a function of entropy $S$ for different values of $\alpha$ and fixed parameters $b=0.1$, $N=0.02$, $\omega=-2/3$, and $P=0.001$. The divergence points correspond to phase transitions identified via the specific heat analysis.}
	\label{fig:RR_vs_S}
\end{figure}

The analysis reveals that the Ruppeiner curvature scalar diverges exactly at the points where the specific heat changes sign, thereby confirming the close correspondence between the geometric description of the thermodynamic state space and the conventional stability analysis. Away from these critical points, both the sign and the magnitude of $R_{R}$ are found to be strongly affected by the cloud of strings parameter $\alpha$ and the quintessential-like fluid parameter $N$, demonstrating that these external fields have a direct impact on the effective microscopic interactions of the black hole. Furthermore, in the regime of large horizon radii, or equivalently large entropy $S$, the curvature scalar approaches zero, indicating that the microscopic structure of the black hole tends toward an ideal-gas-like behavior in this limit (see Fig. \ref{fig:RR_vs_S}).

This geometric perspective thus reinforces the phase structure obtained earlier and offers an interpretation in terms of underlying microphysical interactions. Moreover, it highlights how external matter fields such as a cloud of strings and quintessence-like fluid can imprint distinct geometric signatures in the thermodynamic state space of the black hole.

\section{Inversion Temperature of AdS BH with CoS and QF}\label{S5}

The BH mass $M$ in Eq.\,(\ref{bb1}) can be expressed in terms of the entropy of the BH whose relation to the area of the horizon $A$ is given by $S=A/4=\pi\,r^2_{+}$. Thus, in terms of entropy, Eq.~(\ref{bb1}) simplifies to
\begin{equation}
M =\frac{1}{2\,\sqrt{\pi}}\,\left[\sqrt{S}+\frac{\pi\,\lvert \alpha\rvert\, b^2}{\sqrt{S}}\,{}_2F_1\left(-\frac{1}{2},-\frac{1}{4},\frac{3}{4},-\frac{S^2}{\pi^2\,b^4}\right)-\mathrm{N}\,S^{-3\,w/2}\,\pi^{(3\,w+1)/2}+\frac{8}{3}\,P\,S^{3/2}\right].\label{dd3}
\end{equation}

\begin{figure}[ht!]
    \centering
    \includegraphics[width=0.46\linewidth]{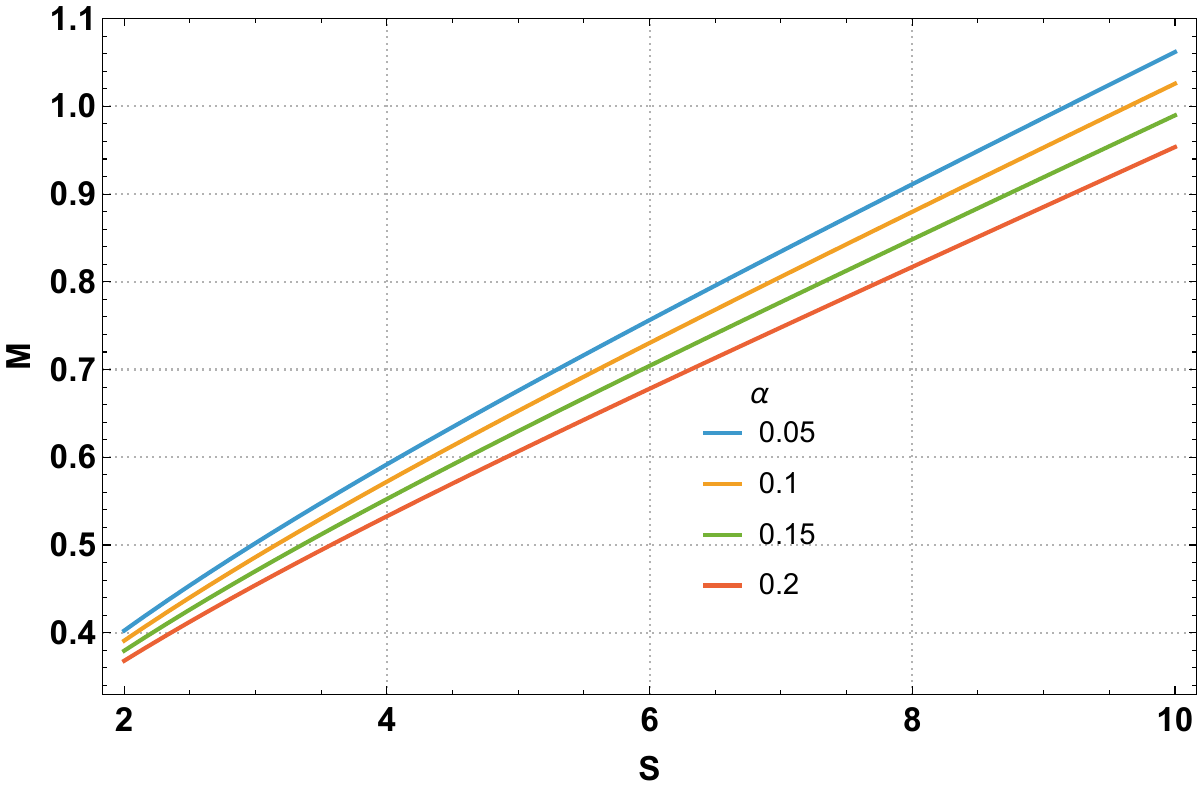}\quad\quad
    \includegraphics[width=0.5\linewidth]{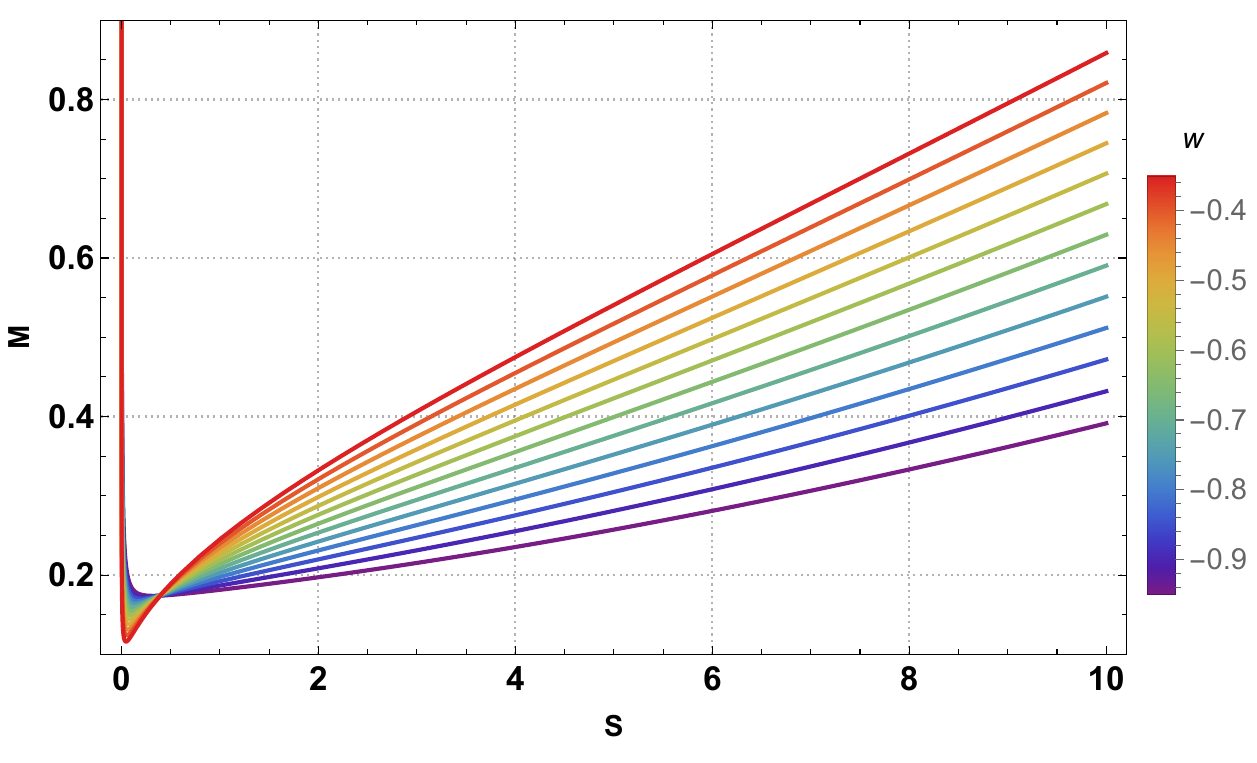}\\
    (i) $b=0.2,\,w=-2/3$ \hspace{6cm} (ii) $\alpha=0.15,\,b=0.2$
    \caption{\footnotesize Behavior of the BH mass $M$ given in Eq. (\ref{dd3}) as a function of entropy for different values of CoS parameter $\alpha$ and state parameter $w$. Here, we set $\mathrm{N}=0.01,\,P=0.01$.}
    \label{fig:BH-mass}
\end{figure}

In Fig. (\ref{fig:BH-mass}), we investigate the behavior of the black hole mass $M$, as defined in Eq.~~(22), by plotting it as a function of the entropy. This analysis is conducted for various values of the (CoS) parameter $\alpha$ and the equation of state parameter $\omega$, while keeping the parameters $N = 0.01$ and $P = 0.01$ fixed throughout. The figure illustrates how modifications in $\alpha$ and $\omega$ influence the thermodynamic mass of the black hole, thereby offering valuable insight into how these parameters affect the underlying gravitational and thermodynamic structure of the system. Specifically, variations in $\alpha$ reflect the influence of the modified gravity framework, whereas changes in $\omega$ encode the effects of different types of matter or energy content. This analysis highlights the intricate dependence of black hole mass on entropy under nontrivial thermodynamic and gravitational corrections, which is essential for a deeper understanding of black hole microphysics in extended theories of gravity.
 
The BH mass of Eq. (\ref{dd3}) can be written as $M=M(S,|\alpha|,b,\mathrm{N})$. Considering the parameters related to the cloud of strings and the quintessence, respectively \((\alpha,b)\) and $\mathrm{N}$, are extensive thermodynamic parameters. As a consequence , the first law of BH thermodynamics must be modified to
\begin{equation}
    dM=T_H\,dS+V\,dP+\mathcal{A}_H\,d|\alpha|+\mathcal{B}_H\,db+\mathcal{C}_H\,d\mathrm{N},\label{mass1}
\end{equation}
where $\mathcal{A}$ and $\mathcal{B}$ are the intensive thermodynamic variables conjugate to the CoS parameters $(|\alpha|,b)$ and $\mathcal{C}$ is another intensive variables related to $\mathrm{N}$. Moreover $T$ is the temperature, $V$ is the volume of the BH. All these variables can be calculated as,
\begin{align}
    T_H&=\left(\frac{\partial M}{\partial S}\right)_{P,|\alpha|,b,\mathrm{N}},\label{mass2}\\
    V&=\left(\frac{\partial M}{\partial P}\right)_{S,|\alpha|,b,\mathrm{N}}=\frac{4}{3}\,S\,\sqrt{S/\pi},\label{mass3}\\
    \mathcal{A}_H&=\left(\frac{\partial M}{\partial |\alpha|}\right)_{S,P,b,\mathrm{N}}=\frac{b^2}{2}\,\sqrt{\frac{\pi}{S}}\,{}_2F_1\left(-\frac{1}{2},-\frac{1}{4},\frac{3}{4},-\frac{S^2}{\pi^2\,b^4}\right),\label{mass4}\\
    \mathcal{B}_H&=\left(\frac{\partial M}{\partial b}\right)_{S,P\,|\alpha|,\mathrm{N}}=\sqrt{\pi/S}\,\lvert \alpha\rvert\,\left[b\,{}_2F_1\left(-\frac{1}{2},-\frac{1}{4},\frac{3}{4},-\frac{S^2}{\pi^2\,b^4}\right)+\frac{S^2}{3\,\pi^2\,b^3}\,{}_2F_1\left( \frac{1}{2}, \frac{3}{4}; \frac{7}{4}; -\frac{S^2}{\pi^2\, b^4} \right)\right],\label{mass5}\\
    \mathcal{C}_H&=\left(\frac{\partial M}{\partial \mathrm{N}}\right)_{S,P,|\alpha|,b}=-\frac{1}{2}\,S^{-3\,w/2}\,\pi^{3\,w/2}=-\frac{1}{2}\,(\pi/S)^{3w/2}.\label{mass6}
\end{align}

\begin{figure}[ht!]
    \centering
    \includegraphics[width=0.47\linewidth]{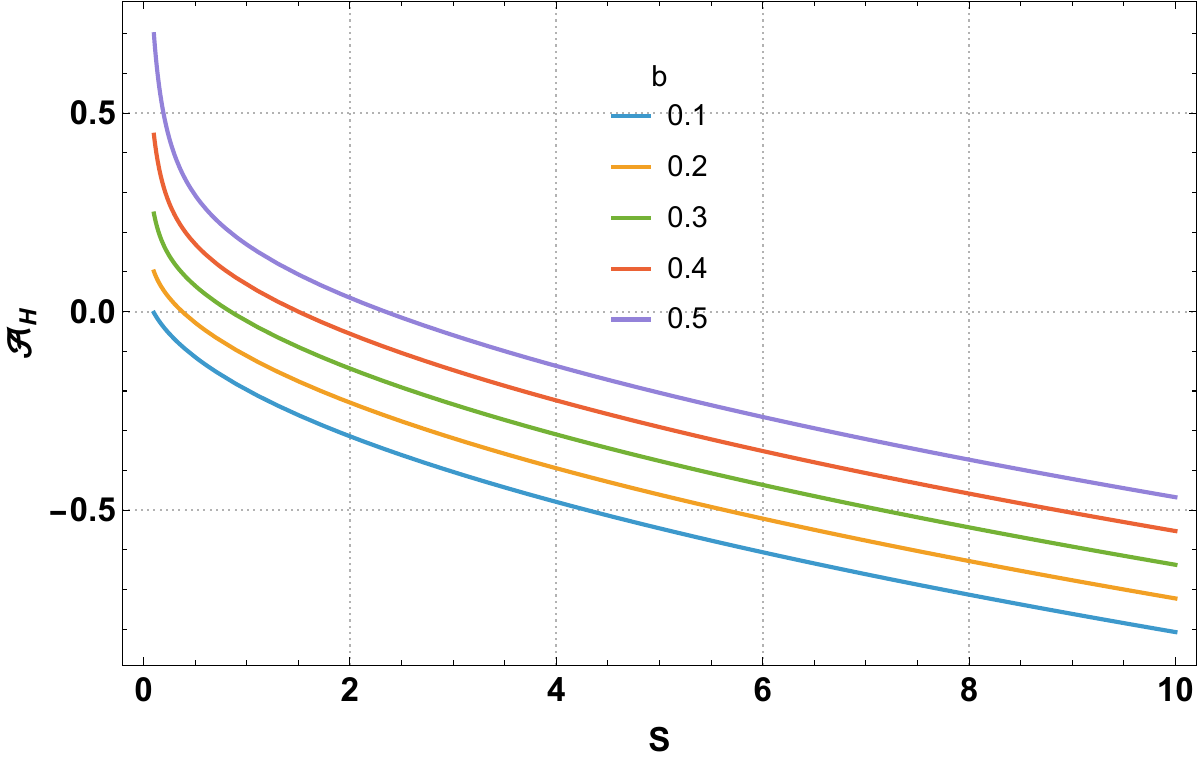}\quad\quad
    \includegraphics[width=0.48\linewidth]{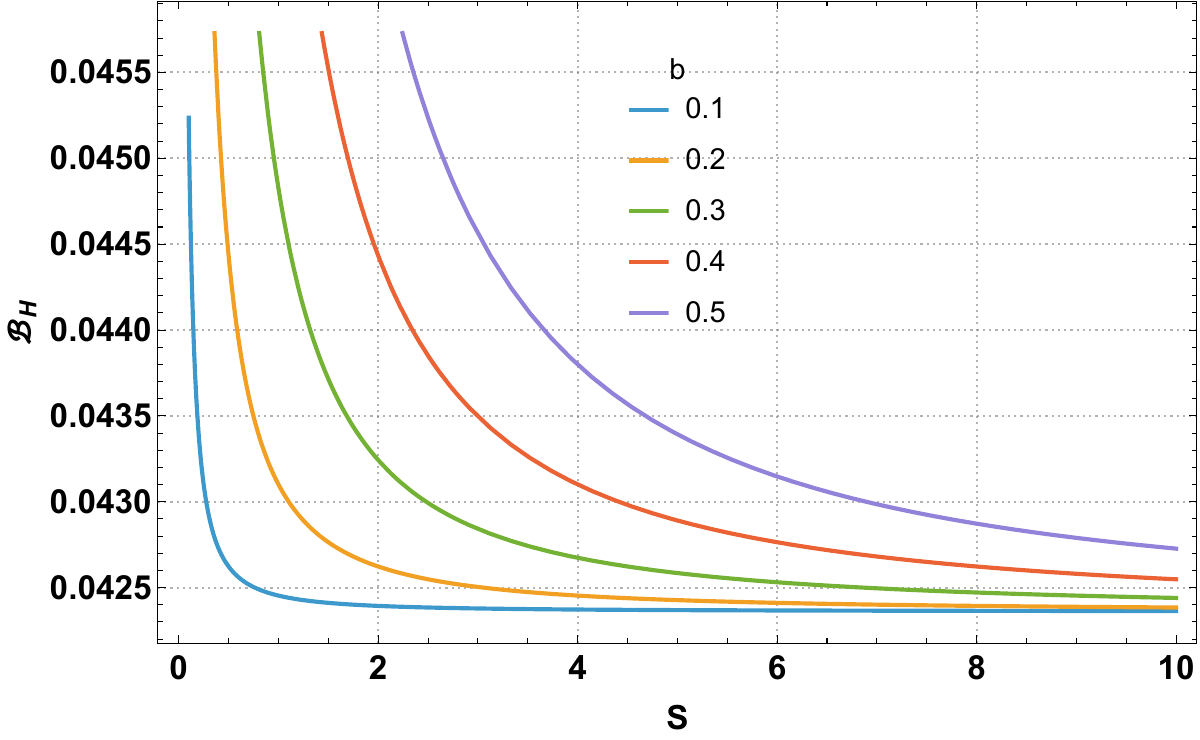}\\
    (i)  \hspace{9cm} (ii) $\alpha=0.05$
    \caption{\footnotesize Behavior of the intensive thermodynamic variables $\mathcal{A}_H$ and $\mathcal{B}_H$ as a function of entropy $S$ for different values of CoS parameter $b$.}
    \label{fig:intensive-variables}
\end{figure}

In Fig.\~(\ref{fig:intensive-variables}), we examine the behavior of the intensive thermodynamic variables $A_H$ and $B_H$ as functions of the entropy $S$, for various values of the (CoS) parameter $b$. This analysis allows us to explore how changes in the parameter $b$, which encapsulates the deviation from standard thermodynamic behavior due to modifications in the underlying gravitational theory or matter content, affect the intensive properties of the black hole system. The figure reveals that both $A_H$ and $B_H$ exhibit distinct and sensitive responses to variations in entropy depending on the chosen value of $b$, indicating that the thermodynamic phase structure is deeply influenced by the specific CoS regime considered. Such an investigation provides a deeper understanding of how intensive variables evolve in extended thermodynamic settings and offers important clues about the microphysical interpretation of black hole thermodynamics in modified gravity scenarios.

Let us now calculate the temperature $T_{H}$ using Eq. (\ref{dd3}). By definition in Eq. (\ref{mass2}), we find the temperature as, 
\begin{align}\label{dd5}
T_H=\frac{1}{4\sqrt{\pi\,S}}\,\Bigg[1-|\alpha|\, b^2\,\left\{
\frac{\pi}{S}\,{}_2F_1\left(-\tfrac{1}{2}, -\tfrac{1}{4}, \tfrac{3}{4}, -\tfrac{S^2}{\pi^2\, b^4} \right)
+ \frac{2\,S}{3\,\pi\,b^4}\,{}_2F_1\left(\tfrac{1}{2}, \tfrac{3}{4}, \tfrac{7}{4}, -\tfrac{S^2}{\pi^2\, b^4} \right)
\right\}+3\,w\, \mathrm{N}\,(\pi/S)^{\tfrac{3\,w+1}{2}}+ 8\,P\,S\Bigg].
\end{align}
We can see that the temperature $T_H$ when express in terms of horizon is the same expression as the Hawking temperature $T$ obtained in Eq.(\ref{bb2}). 

\begin{figure}[ht!]
    \centering
    \includegraphics[width=0.485\linewidth]{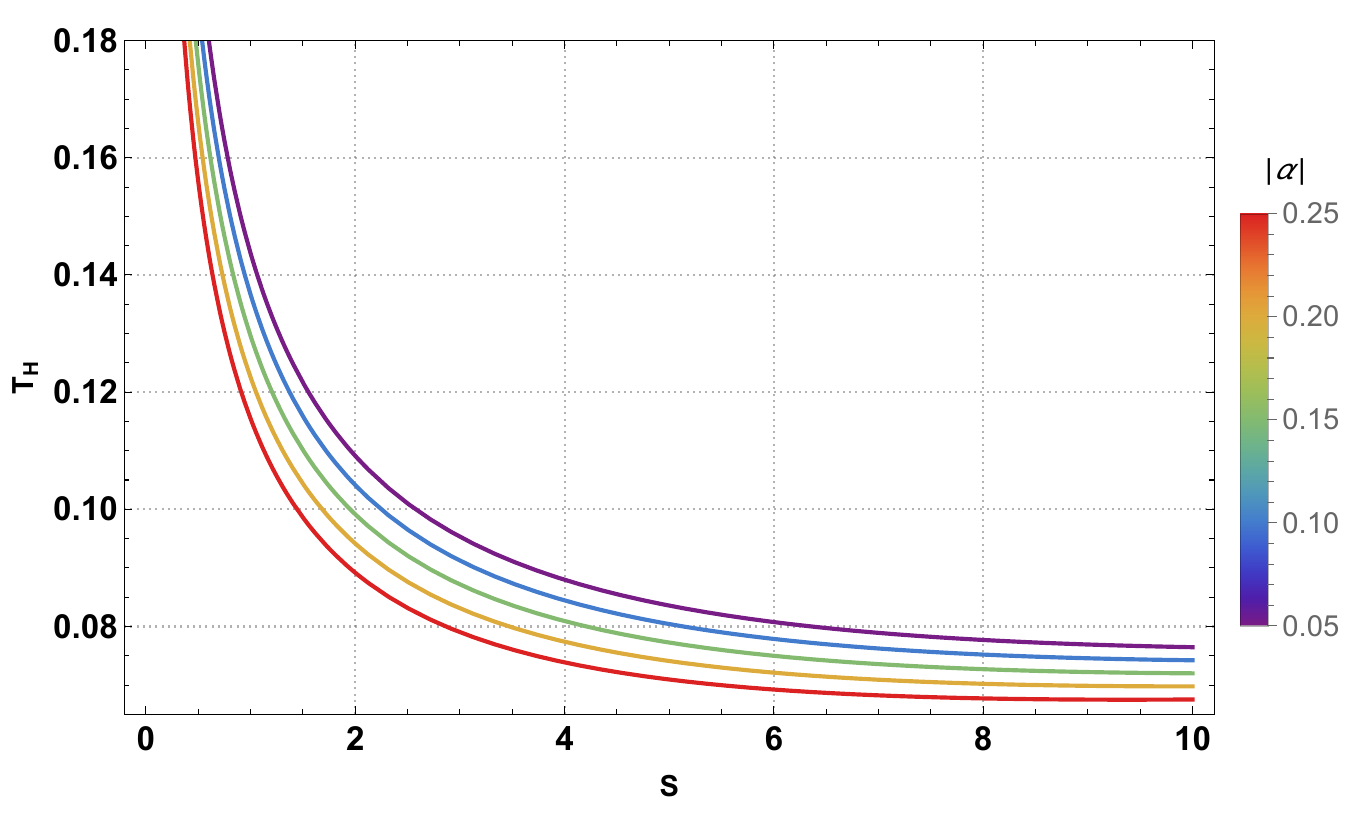}\quad
    \includegraphics[width=0.485\linewidth]{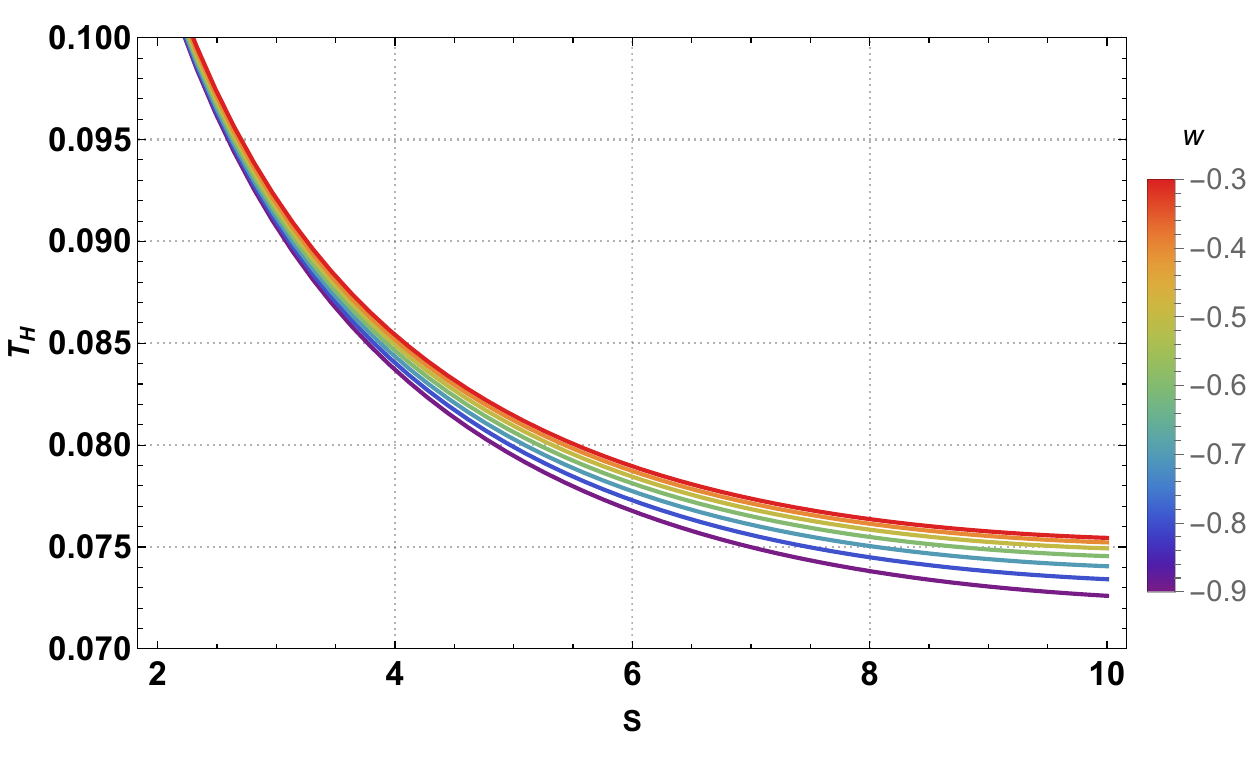}\\
    (i) $b=0.1,\,w=-2/3$  \hspace{6cm} (ii) $\alpha=0.1,\,b=0.1$
    \caption{\footnotesize Behavior of the temeprature $T_H$ given in Eq. (\ref{dd5}) as a function of entropy for different values of CoS parameter $\alpha$ and state parameter $w$. Here, we set $\mathrm{N}=0.01,\,P=0.01$.}
    \label{fig:temperature-entropy}
\end{figure}

In Fig.~~(\ref{fig:temperature-entropy}), we analyze the behavior of the Hawking temperature $T_H$, as defined in Eq.~~(29), plotted as a function of the entropy $S$ for different values of the (CoS) parameter $\alpha$ and the equation of state parameter $\omega$. For this analysis, the parameters $N = 0.01$ and $P = 0.01$ are held fixed to isolate the effects of $\alpha$ and $\omega$ on the thermodynamic profile of the system. The figure clearly demonstrates how variations in both $\alpha$ and $\omega$ influence the temperature behavior, thereby reflecting the sensitivity of black hole thermodynamics to modifications in the gravitational sector and matter content. In particular, changes in $\alpha$ can be associated with deviations from classical general relativity, while different values of $\omega$ correspond to various types of energy-matter configurations. The resulting temperature-entropy relationship provides valuable insights into the thermal stability and phase structure of black holes within the framework of modified gravity theories and extended thermodynamics.

In addition, the thermodynamic volume of the BH in terms of horizon can be computed as
\begin{equation}\label{dd6}
V = \left( \frac{\partial M}{\partial P} \right)= \frac{4\pi r_+^3}{3}\Rightarrow r_{+}=(3V/4\pi)^{1/3}.
\end{equation}
Hence, the Hawking temperature given in Eq.~(\ref{bb2}) as a function of the BH volume $V$ reads as,
\begin{align}\label{dd7}
T&= \frac{1}{4\pi} \Bigg[\left( \frac{4\pi}{3V} \right)^{1/3}- |\alpha|\, b^2 \left\{
\left( \frac{4\pi}{3V} \right)\,{}_2F_1\left(-\tfrac{1}{2}, -\tfrac{1}{4}, \tfrac{3}{4}, 
- \left( \frac{3\,V}{4\,\pi\,b^3} \right)^{4/3} \right)
+ \frac{2}{3\,b^4} \left( \frac{3V}{4\pi} \right)^{1/3}\,{}_2F_1\left(\tfrac{1}{2}, \tfrac{3}{4}, \tfrac{7}{4}, 
- \left( \frac{3V}{4\pi b^3} \right)^{4/3} \right)
\right\} \nonumber\\
&+ 3w\,\mathrm{N}\,\left( \frac{4\pi}{3V} \right)^{(3w+2)/3}
+ 8\pi P \left( \frac{3V}{4\pi} \right)^{1/3}
\Bigg]
\end{align}

In order to get the inversion temperature, one can use its definition \cite{BZM,JT10a,JTa,JT}, so that
\begin{equation}\label{dd8}
T_i = V \left( \frac{\partial T}{\partial V} \right)_{P,|\alpha|,b,\mathrm{N}}.
\end{equation}
By using Eq.\,(\ref{dd7}) and computing the inversion temperature, one arrives
\begin{align}\label{dd9}
T_i&=-\frac{1}{9} \left( \frac{3V}{4\pi} \right)^{-\frac{1}{3}}+\frac{|\alpha| b^{2}}{3} \left( \frac{3V}{4\pi} \right)^{-1}\,{}_2F_1\left( -\frac{1}{2}, -\frac{1}{4}, \frac{3}{4}, -\frac{\left( \frac{3V}{4\pi} \right)^{\frac{4}{3}}}{b^{4}} \right)+\frac{4|\alpha|}{21 b^{6}}\,\left(\frac{3V}{4\pi} \right)^{\frac{5}{3}}\,{}_2F_1\left( \frac{3}{2}, \frac{7}{4}, \frac{11}{4}, -\frac{\left( \frac{3V}{4\pi} \right)^{\frac{4}{3}}}{b^{4}} \right)\nonumber\\
&- \frac{1}{3} N w (3w+2) \left( \frac{3V}{4\pi} \right)^{-\frac{3w+2}{3}}+\frac{8\pi P}{9} \left( \frac{3V}{4\pi} \right)^{\frac{1}{3}}
\end{align}

In terms of horizon radius \(r_+\), the inversion temperature canbe rewritten as,
\begin{align}\label{dd10}
T_i &= - \frac{1}{9 r_+}
+ \frac{|\alpha| b^{2}}{3 r_+^{3}}\,{}_2F_1\left(-\frac{1}{2}, -\frac{1}{4}, \frac{3}{4}, -\frac{r_+^{4}}{b^{4}}\right)+ \frac{4 |\alpha| r_+^{5}}{21 b^{6}}\,{}_2F_1\left(\frac{3}{2}, \frac{7}{4}, \frac{11}{4}, -\frac{r_+^{4}}{b^{4}}\right)- \frac{1}{3}\,\mathrm{N}\, w\, (3w+2)\, r_+^{-(3w+2)}
+ \frac{8 \pi P r_+}{9}
\end{align}

From expressions (\ref{dd9}) or (\ref{dd10}), we observe that the inversion temperature is modified by the CoS parameters $(\alpha,\,b)$ and the QF parameters \((\mathrm{N},\,w)\).

Finally, subtracting Eqs.\,(\ref{dd10}) and (\ref{bb2}), one gets
\begin{align}\label{dd11}
\Delta T &=T_i - T = -\left( \frac{1}{9} + \frac{1}{4\pi} \right) \frac{1}{r_+} + |\alpha|\,b^{2}\, \left( \frac{1}{3} + \frac{1}{4\pi} \right)\, \frac{1}{r_+^{3}} \, {}_2F_1\left(-\frac{1}{2}, -\frac{1}{4}, \frac{3}{4}, -\frac{r_+^{4}}{b^{4}} \right)+ \frac{|\alpha|\, r_+}{6\, \pi\, b^{2}} \, {}_2F_1\left( \frac{1}{2}, \frac{3}{4}, \frac{7}{4}, -\frac{r_+^{4}}{b^{4}} \right)\nonumber\\
&+ \frac{4\, |\alpha|\, r_+^{5}}{21\, b^{6}} \, {}_2F_1\left( \frac{3}{2}, \frac{7}{4}, \frac{11}{4}, -\frac{r_+^{4}}{b^{4}} \right)+ \mathrm{N} r_+^{-(3\,w+2)} \left[ -\frac{1}{3}\, w \,(3\,w + 2) - \frac{3\,w}{4\,\pi} \right]+ P\, r_+ \left( \frac{8 \pi}{9} - 2 \right)
\end{align}

For a particular state parameter, $w=-2/3$, we find
\begin{align}\label{dd12}
\Delta T &=-\left( \frac{1}{9} + \frac{1}{4\pi} \right) \frac{1}{r_+} + |\alpha|\,b^{2}\, \left( \frac{1}{3} + \frac{1}{4\pi} \right)\, \frac{1}{r_+^{3}} \, {}_2F_1\left(-\frac{1}{2}, -\frac{1}{4}, \frac{3}{4}, -\frac{r_+^{4}}{b^{4}} \right)+ \frac{|\alpha|\, r_+}{6\, \pi\, b^{2}} \, {}_2F_1\left( \frac{1}{2}, \frac{3}{4}, \frac{7}{4}, -\frac{r_+^{4}}{b^{4}} \right)\nonumber\\
&+ \frac{4\, |\alpha|\, r_+^{5}}{21\, b^{6}} \, {}_2F_1\left( \frac{3}{2}, \frac{7}{4}, \frac{11}{4}, -\frac{r_+^{4}}{b^{4}} \right)+ \mathrm{N}\,\frac{2}{4\,\pi}+ P\, r_+ \left( \frac{8 \pi}{9} - 2 \right)
\end{align}

From the above analysis, it is evident that the Hawking or BH temperature exhibits a form significantly different from that of the inversion temperature, which arises when the Joule-Thomson expansion coefficient of the system vanishes. These two are generally not equal and describe very different physical processes: 
\begin{itemize}
    \item The Hawking temperature is associated with the thermal radiation emitted by a BH due to quantum effects near the event horizon. In contrast, the inversion temperature arises from the Joule-Thomson expansion in extended BH thermodynamics, defined as the temperature at which the Joule-Thomson coefficient vanishes.

    \item The Hawking temperature governs the rate of BH evaporation and is directly related to the surface gravity at the event horizon. On the other hand, the inversion temperature marks the critical point at which the behavior of the BH during an adiabatic expansion at constant enthalpy changes from cooling to heating or vice versa.

    \item While the Hawking temperature is fundamental for understanding BH radiation and quantum field effects in curved space-time, the inversion temperature characterizes the BH as a thermodynamic system in an extended phase space, where the cosmological constant is interpreted as pressure and mass as enthalpy. It reflects a behavior analogous to that of real gases in classical Joule-Thomson expansion processes.
\end{itemize}

\section{Joule-Thomson Expansion of AdS BH with CoS and QF} \label{S6}

In this section, we will conduct a detailed investigation of the J-T expansion as it applies to the Schwarzschild-AdS black holes with cloud of strings and quintessential-like fluid. The J-T expansion, a classical thermodynamic process, provides insight into the temperature behavior of a system undergoing adiabatic expansion. To draw a parallel with conventional thermodynamics, the J-T expansion typically refers to a throttling process involving a non-ideal fluid, such as a gas or liquid. In this process, the fluid is forced through a valve or porous plug, moving from a region of higher pressure to one of lower pressure. Importantly, this expansion occurs without heat exchange with the surroundings, hence it is adiabatic. While the process is inherently irreversible, the enthalpy of the fluid remains constant throughout, making it an isenthalpic process. This key feature allows for the study of temperature changes that result solely from variations in pressure and volume under constant enthalpy conditions. The change in temperature during the J-T expansion is quantified by the J-T coefficient, denoted by $\mu$, which is defined as the rate of change of temperature $T$ with respect to pressure $P$ at constant enthalpy $H$ \cite{BZM,JT10a,JTa,JT}:
\begin{equation}\label{K1}
\mu = \left( \frac{\partial T}{\partial P} \right)_H = \frac{1}{C_P} \left[ T \left( \frac{\partial V}{\partial T} \right)_P - V \right].
\end{equation}

\begin{figure}[ht!]
 \centering
 \includegraphics[width=0.48\textwidth]{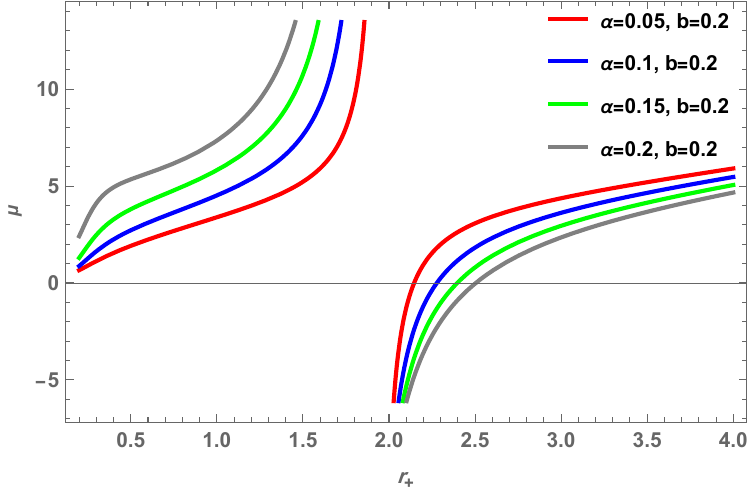}\quad
 \includegraphics[width=0.48\textwidth]{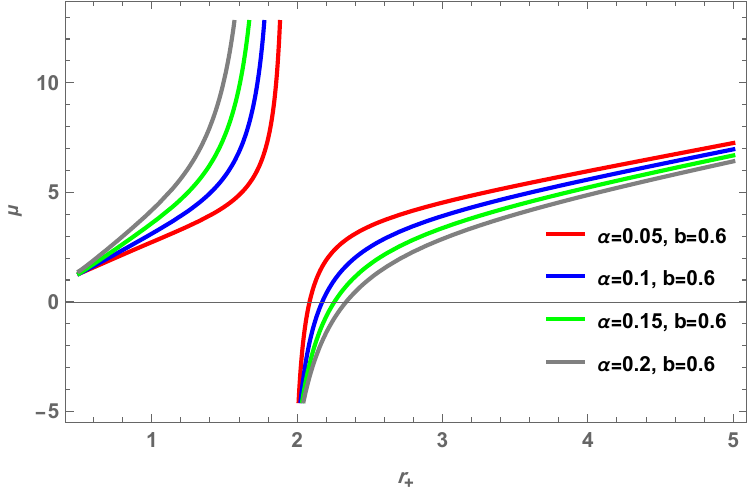}\\
 (i) \hspace{8cm} (ii)\\
 \includegraphics[width=0.48\textwidth]{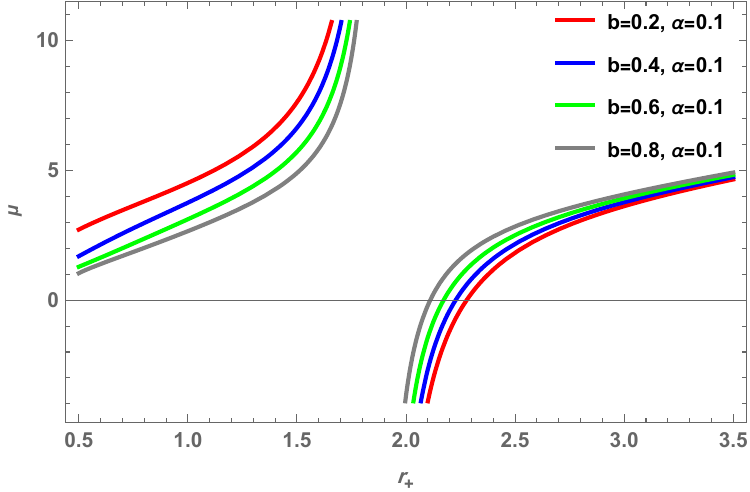}\quad
 \includegraphics[width=0.48\textwidth]{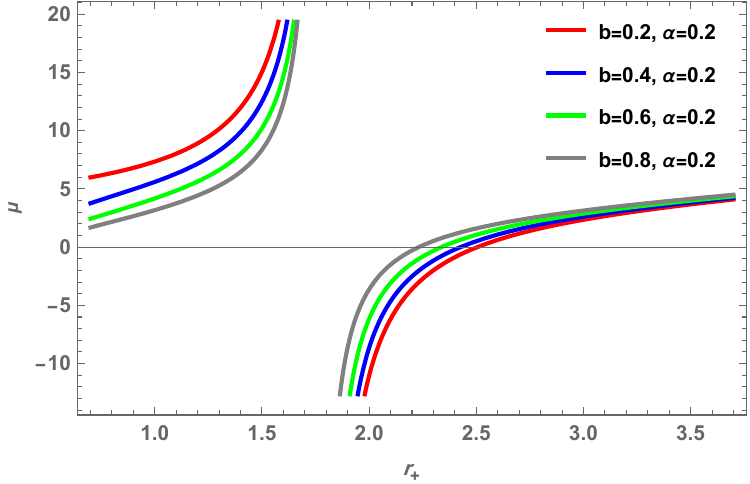}\\
 (iii) \hspace{8cm} (iv)
  \caption{\footnotesize Profile of J-T coefficient as a function of entropy for fixed parameters $N = 0.01$, P=0.01 and $w = -\frac{2}{3}$, plotted for various values of the parameters $\alpha$ and $b$. The figure illustrates how changes in $\alpha$ and $b$ influence the thermodynamic behavior and inversion points of the J-T expansion}
 \label{N6}
 \end{figure}

Here, $C_P$ represents the heat capacity at constant pressure per unit horizon volume of the BH system. The subscript $H$ indicates that the partial derivative is evaluated along an isenthalpic process. In the context of BH thermodynamics within extended phase space, the enthalpy is identified with the BH, $M$. This relation is crucial because it links thermodynamic variables-temperature $T$, pressure $P$, and volume $V$, allowing us to determine whether the system cools or heats during expansion. A positive JT coefficient ($\mu > 0$) indicates cooling upon expansion, while a negative value ($\mu < 0$) corresponds to heating (Fig. (\ref{N6})). 

By applying this formalism to Schwarzschild-AdS BH with CoS and QF, we aim to gain a deeper understanding of their thermodynamic behavior under J-T expansion. So with respect to Eq. (\ref{K1}), we obtain J-T coefficient as,
\begin{equation}
    \mu=\frac{\mathrm{A}}{\mathrm{B}},\label{K3}
\end{equation}
where we have defined 
\begin{equation}\label{K2}
\begin{split}
&\mathrm{A} = \frac{
2 r_+ \left( 
|\alpha| \left( 
-14 b^4 r_+^4 \, {}_2F_1\left(\tfrac{1}{2},\tfrac{3}{4};\tfrac{7}{4};-\tfrac{r_+^4}{b^4}\right)
+18 r_+^7 \, {}_2F_1\left(\tfrac{3}{2},\tfrac{7}{4};\tfrac{11}{4};-\tfrac{r_+^4}{b^4}\right)
-63 b^8 \, {}_2F_1\left(-\tfrac{1}{2},-\tfrac{1}{4};\tfrac{3}{4};-\tfrac{r_+^4}{b^4}\right)
\right)
+21 b^6 r_+^2 (1 - 8 \pi P r_+^2)
\right)
}{
21 b^4 \left(
|\alpha| \left( 
3 b^4 \, {}_2F_1\left(-\tfrac{1}{2}, -\tfrac{1}{4}; \tfrac{3}{4}; -\tfrac{r_+^4}{b^4}\right) 
+ 2 r_+^3 \, {}_2F_1\left(\tfrac{1}{2}, \tfrac{3}{4}; \tfrac{7}{4}; -\tfrac{r_+^4}{b^4} \right)
\right)
-3 b^2 r_+^2 (-2 N r_+ + 8 \pi P r_+^2 + 1)
\right)
} \\
&\mathrm{B}= \frac{
b^2 r_+ \sqrt{ \tfrac{r_+^4}{b^4} + 1 } 
\left( 3 N w (3 w + 5) + 4 (4 \pi P r_+^2 + 1) r_+^{3w + 1} \right)
- 2 |\alpha| (3 b^4 + 2 r_+^4) r_+^{3w}
}{
|\alpha| (3 b^4 + r_+^4) r_+^{3w}
+ b^2 r_+ \sqrt{ \tfrac{r_+^4}{b^4} + 1 } 
\left( (8 \pi P r_+^2 - 1) r_+^{3w + 1} - 3 N w (3w + 2) \right)
}.
\end{split}
\end{equation}

\section{Conclusion} \label{S7}

In this paper, the analysis of the Schwarzschild-AdS BH surrounded by a cloud of strings and a quintessential-like fluid reveals that the thermodynamic behavior of the system is profoundly influenced by the parameters associated with the external matter fields and the AdS background. By interpreting the BH mass as enthalpy within the extended phase space formalism and associating the cosmological constant with thermodynamic pressure, we derived explicit expressions for the mass, Hawking temperature, Gibbs free energy, and internal energy, all of which exhibit nontrivial dependence on the cloud of strings parameters $(\alpha, b)$, the quintessence parameters $(N, w)$, and the pressure $P$. Also, we know that the zero-temperature points corresponding to divergence in the Joule–Thomson coefficient, marking critical boundaries between heating and cooling phases in expansion or compression processes. For the horizon radii, the thermal curves reveal strong sensitivity to the microphysical parameters, indicating that the presence of the CoS and QF significantly alters near-horizon thermodynamics, while for large BHs these effects become subdominant, leading to a universal asymptotic thermal behavior. The Gibbs free energy analysis, as its dependence on $r_+$ illustrates potential phase transitions characterized in $F(r_+)$, indicative of first-order or higher-order critical phenomena. In this context, the study of thermodynamics suggests the presence of phase transitions between distinct equilibrium configurations. Also, the internal energy structure underscores the role of external fields in shaping the BH’s energetics, give more explanation of the coupling between gravitational, matter, and cosmological contributions that generates a thermodynamic phase space. 

The stability analysis of Schwarzschild-AdS BHs in the presence of a cloud of strings and a quintessential-like fluid, based on the behavior of the specific heat $C_p$, reveals a rich thermodynamic structure influenced by the parameters $(\alpha, b)$ and $(\mathrm{N}, w)$. The explicit expressions derived from the Hawking temperature and entropy demonstrate that divergences in $C_p$ occur at certain critical horizon radii, signaling second-order phase transitions where the BH undergoes qualitative thermodynamic changes. Also, cases with $C_p > 0$ correspond to thermally stable configurations capable of sustaining equilibrium under small perturbations, whereas $C_p < 0$ indicates instability, a behavior common in BH thermodynamics. 

The thermodynamic geometry results show that the Ruppeiner curvature scalar $R_{R}$ for the Schwarzschild-AdS BH with a cloud of strings and a quintessential-like fluid provides a clear geometric signature of its phase structure and microscopic behavior. The calculated $R_{R}$ diverges exactly at the points where the specific heat changes sign, confirming that phase transitions identified through standard thermodynamic analysis have a direct correspondence in the geometric description of the state space. The sign of $R_{R}$ distinguishes between dominant repulsive ($R_{R} > 0$) or attractive ($R_{R} < 0$) microscopic interactions, and both its sign and magnitude are found to be strongly affected by the parameters $\alpha$ and $N$, representing the string cloud and quintessence-like fluid respectively. This demonstrates that external matter fields can significantly modify the BH’s effective microscopic interaction landscape. In the limit of large entropy, $R_{R}$ approaches zero, indicating that the system’s microstructure tends toward ideal gas–like behavior, reinforcing the view that the geometric approach not only complements but deepens the thermodynamic stability analysis.

 The results demonstrate that the inversion temperature of a Schwarzschild-AdS BH is significantly influenced by the presence of a cloud of strings (CoS) and a quintessence-like fluid (QF). By extending the first law of BH thermodynamics to include the parameters $(|\alpha|, b)$ for the CoS and $(\mathrm{N}, w)$ for the QF, the derived expressions reveal how these additional components modify both the Hawking temperature and the inversion temperature compared to the standard AdS case. The analysis tests and shows how variations in these parameters shift the inversion curve, altering the critical points at which the BH transitions between heating and cooling during the Joule–Thomson expansion. In particular, the difference $\Delta T$ between the inversion and Hawking temperatures highlights the distinct thermodynamic response induced by CoS and QF effects, with the case $w=-2/3$ providing a simplified example where the modifications become more transparent. This indicates that the coupling of BHs with such external fields can profoundly reshape their thermal phase structure. 

The results of the Joule–Thomson expansion analysis reveal that the Schwarzschild-AdS BH with a cloud of strings and a quintessential-like fluid exhibits a distinct dependence of its cooling–heating behavior on the parameters $\alpha$ and $b$. The explicit form of the J-T coefficient $\mu$ shows that variations in these parameters shift the inversion points, thereby altering the regions where the BH cools ($\mu > 0$) or heats ($\mu < 0$) during isenthalpic expansion. As illustrated , increasing $\alpha$ generally enhances the cooling region, while changes in $b$ influence both the magnitude and location of the inversion temperature. Also, this indicates that the interplay between the string cloud density, the nonlinear electromagnetic field strength, and the quintessential fluid significantly affects the thermodynamic response of the system, offering valuable insights into the control of phase behavior in extended BH thermodynamics.

{\footnotesize

\section*{Acknowledgments}
F.A. acknowledges the Inter University Centre for Astronomy and Astrophysics (IUCAA), Pune, India, for granting a visiting associateship.

\section*{Data Availability Statement}

This manuscript has no associated data. [Author’s comment: Data sharing not applicable to this article as no datasets were generated or analysed during the current study.]

\section*{Code Availability Statement}

This manuscript has no associated code/software. [Author’s comment: Code/Software sharing not applicable to this article as no code/software was generated or analyzed during the current study.]
}

\end{document}